\documentclass[letterpaper]{JHEP3}
\usepackage{amssymb}

\usepackage{epsfig}

\newcommand{\beq}{\begin{equation}}
\newcommand{\eeq}{\end{equation}}
\newcommand{\beqa}{\begin{eqnarray}}
\newcommand{\eeqa}{\end{eqnarray}}
\newcommand{\beqar}{\begin{eqnarray*}}
\newcommand{\eeqar}{\end{eqnarray*}}

\newcommand{\norm}[1]{\raise.3ex\hbox{:}#1\raise.3ex\hbox{:}}

\newcommand{\bm}{\begin{multiline}}
\newcommand{\beqs}{\begin{eqnarray}}
\newcommand{\eeqs}{\end{eqnarray}}

\abstract{ We construct new explicit solutions of general
relativity from double analytic continuations of Taub-NUT
spacetimes. This generalizes previous studies of $4$-dimensional
nutty bubbles. One $5$-dimensional locally asymptotically AdS
solution in particular has a special conformal boundary structure
of $AdS_3\times S^1$. We compute its boundary stress tensor and
relate it to the properties of the dual field theory.
Interestingly enough, we also find consistent $6$-dimensional
bubble solutions that have only one timelike direction. The
existence of such spacetimes with non-trivial topology is closely
related to the existence of the Taub-NUT(-AdS) solutions with more
than one NUT charge. Finally, we begin an investigation of
generating new solutions from Taub-NUT spacetimes and nuttier
bubbles. Using the so-called Hopf duality, we provide new explicit
time-dependent backgrounds in six dimensions. } \preprint{hep-th/0508162}

\title{Nuttier Bubbles}
\author{Dumitru Astefanesei,$^1$\thanks{%
E-mail: \texttt{dastef@mri.ernet.in}} \ Robert B. Mann$^{2}$\thanks{%
E-mail: \texttt{mann@avatar.uwaterloo.ca}} \ and Cristian Stelea$^3$\thanks{%
E-mail: \texttt{cistelea@uwaterloo.ca}} \\
$^{1}$Harish-Chandra Research Institute, Chhatanag Road, Jhusi, Allahabad
211019, India\\
$^{1,2}$Perimeter Institute for Theoretical Physics, Ontario N2J 2W9, Canada%
\\
$^{2,3}$Department of Physics, University of Waterloo Waterloo, Ontario N2L
3G1, Canada}

\begin{document}




\section{Introduction}

Many important problems in physics, such as cosmological evolution or black
hole evaporation, involve time in an essential way. Therefore, a key problem
in string theory is understanding its behaviour in time-dependent
backgrounds. In order to carry out this investigation one needs to construct
simple enough time dependent-solutions that would provide consistent
test-beds on which one could try to address these problems.

`Bubbles of nothing' are smooth, time-dependent\footnote{%
There are also time-independent bubbles. Properties of bubbles of nothing in
different situations are studied in \cite%
{{witten},{Aharony:2002cx},{Birmingham:2002st},{Balasubramanian:2002am},
{nuttybubbles},{Biswas:2004xc},{Astefanesei:2005eq},{eugen},{cai},{radu}}.} vacuum
solutions of Einstein's equations and so are consistent backgrounds for
string theory, at least at leading order. The characteristic feature of such
a solution is that it has a (minimal) area with \textit{no} space inside.
For example, a bubble solution can be obtained from a four-dimensional
static, spherically symmetric black hole by a double analytic continuation
in the time coordinate and some other combination of coordinates on the $%
S^{2}$-section. This way, the sphere is effectively changed into a
two-dimensional de Sitter spacetime.

The first example was provided by Witten \cite{witten} as the endstate of
the decay of the Kaluza-Klein (KK) vacuum. Balasubramanian \textit{et at.} %
\cite{Balasubramanian:2005bg} found a similar process in Anti-de Sitter
(AdS) spacetime. That is, a certain orbifold in AdS (analogue of the flat
space KK vacuum) decays via a bubble of nothing. This opens the possibility
that highly non-perturbative processes in gravity might be described (via
AdS/CFT correspondence \cite{maldacena}) as barrier penetration in a dual
field theory effective potential.

Recently there has been renewed interest in bubbles of nothing since it was
pointed out that they provide new endpoints for Hawking evaporation \cite%
{Horowitz:2005vp}. Closed string tachyon condensation is at the basis of a
topology changing transition from black strings to bubbles of nothing.

In this paper we will extend the work of \cite{nuttybubbles} in higher
dimensions, in which so-called `nutty bubbles' --- time-dependent
backgrounds obtained by double analytic continuations of the
coordinates/parameters of (locally asymptotically flat AdS) NUT-charged
solutions --- were obtained.\footnote{%
Intuitively, the NUT charge corresponds to a `magnetic' type of mass. The
Taub-NUT spacetimes and their boundary geometries are relevant for
gauge/gravity dualities \cite{noi} --- see, also, \cite{vazquez} and
references therein for other applications.} The gravitational instantons
associated with NUT-charged spacetimes come in two classes: `nuts' and
`bolts'. In four dimensions the topology of nut is $R^{4}$ and the apparent
singularity at the origin is nothing but a coordinate singularity of the
polar coordinate system --- in this context (AdS) Taub-NUT is an extremal
background. On the other hand, the (AdS) Taub-Bolt geometry should be
thought of as a Euclidean black hole with a NUT charge and non zero
temperature. In certain situations the NUT charge induces an ergoregion into
the bubble spacetime and in other situations it quantitatively modifies the
evolution of the bubble, as when rotation is present.

It was conjectured in \cite{nuttybubbles} that one cannot construct
consistent nutty bubble solutions (with only one timelike direction) in
higher dimensions because (at that time) no $5$ (or higher)-dimensional
NUT-charged solutions with more than one NUT charge were known. Recently
such generalized NUT-charged solutions were obtained \cite%
{csrm,TNletter,Page1} --- from these we are able to provide interesting
time-dependent bubble solutions in higher-dimensions.

Our paper is organized as follows: in the next section we construct
bubble solutions starting from the $5$-dimensional solutions presented in %
\cite{csrm,Page1}. In five dimensions the Taub-NUT solutions have only one
nut parameter. Moreover there is a restriction that connects the value of
the nut parameter to the cosmological constant. The bubbles are obtained by
performing appropriate double analytic continuations of the coordinates.
While our focus is primarily on the asymptotically AdS solutions, we also
provide non-trivial bubble solutions that are asymptotically dS as well as a 
$5$-dimensional non-asymptotically flat solution. Remarkably, we find a
locally asymptotically AdS solution with a boundary geometry of $%
AdS_{3}\times S^{1}$. In the Discussion section we calculate its boundary
stress tensor and show that it has two pieces: one that depends on the
parameters of the bubble, and the other one which is universal and is
reproduced by the universal anomaly contribution to the stress tensor of
Yang-Mills theory on $AdS_{3}\times S^{1}$.

In the third section we construct interesting higher dimensional nutty
bubbles from some of the $6$-dimensional Taub-NUT solutions presented in %
\cite{csrm,TNletter}, however we focus on describing in detail only a couple
of representative $6$-dimensional spaces. In contrast to the
lower-dimensional spaces, in higher than $6$-dimensions there can be at
least two independent nut parameters and, quite generically, there exits a
set of constraints that relate the values of these nut parameters to the
cosmological constant. We can analytically continue the nut parameters
independently, as long as we can still satisfy the constraints (or their
analytically continued avatars). We consider first the case when the base
space of the circle fibration characteristic of the Taub-NUT solution is $%
S^{2}\times S^{2}$. In this case there are two independent nut parameters
only if the cosmological constant vanishes. We also treat in Appendix A the
case in which the base space is of the form $M_{1}\times M_{2}$, where the $2
$-dimensional factors $M_{i}$ are distinct (in which case the cosmological
constant is non-vanishing). In six dimensions there exists a different
class of cosmological Taub-NUT solutions, which are characterized by one nut
parameter only. The fibration is constructed over the first $2$-dimensional
factor $M_{1}$ and we consider the warped product with $M_{2}$. The novelty
of this type of solution is that the warp factor depends non-trivially on
the cosmological constant and the NUT charge. The time-dependent bubble
solutions are obtained by double analytic continuations of the coordinates $%
and$ the nut parameter. Finally, we also present a method to generate new
time-dependent solutions by using Hopf-dualities. Essentially, Hopf duality
is a $T$-duality applied along the $U(1)$-fibre characteristic to the
Taub-NUT-like fibrations \cite{Duff1,Duff2,Cvetic,KKM}. We apply this method
to some of our $6$-dimensional bubble solutions to generate new
time-dependent backgrounds. We end our paper with a discussion section. In
Appendix A, we present other examples of $6$-dimensional time-dependent
solutions, constructed starting from the Taub-NUT spaces whose basis are of
the form $S^{2}\times T^{2}$, respectively $T^{2}\times T^{2}$. For convenience, in Appendix B we present a derivation of the Hopf-duality rules.

\section{Five dimensional `nutty' bubbles}

In four dimensions the usual Taub-NUT construction corresponds to a
circle-fibration over a base space that is a two-dimensional Einstein-K$\dot{%
a}$hler manifold. This base space is usually taken to be the sphere $S^{2}$
--- however, it can also be the torus $T^{2}$ or the hyperboloid $H^{2}$. In
five dimensions the corresponding base space is three dimensional and
consequently the above construction is not straightforward. A
five-dimensional Taub-NUT space built as a `partial' fibration over a
two-dimensional Einstein-K\"ahler space was given in \cite{csrm,Page1}%
. However, in five dimensions there is a constraint on the possible values
of the nut charge and the cosmological constant. The effect of this
constraint is such that for a circle fibration over the sphere, $S^2$, the
cosmological constant can take only positive values, for a fibration over
the torus, $T^2$, the cosmological constant must vanish, while in the case
of a fibration over the hyperboloid, $H^2$, the cosmological constant can
have only negative values.\footnote{%
We are considering here the Lorentzian sections of the metric.} It is worth
mentioning that one cannot simultaneously set the nut charge and/or the
cosmological constant to zero --- \textit{i.e.} there is no smooth limit in
which one can obtain five dimensional Minkowski space in this way. However,
in the Discussion section, we will present ways to evade this situation.

\subsection{Nutty bubbles in AdS}

As noted above, the cosmological constant can be negative ($\Lambda =-\frac{6%
}{l^{2}}$) only in the case of a fibration over the hyperboloid $H^{2}$. The
metric is 
\begin{equation}
ds^{2}=-F(r)(dt-2n\cosh \theta d\phi
)^{2}+F^{-1}(r)dr^{2}+(r^{2}+n^{2})(d\theta ^{2}+\sinh ^{2}\theta d\phi
^{2})+r^{2}dy^{2},  \label{56}
\end{equation}%
where 
\begin{equation}
F(r)=\frac{4r^{4}+2l^{2}r^{2}-16ml^{2}}{l^{2}(4r^{2}+l^{2})}.  \label{57}
\end{equation}%
Moreover, there is a constraint on the nut parameter $n^{2}=\frac{l^{2}}{4}$%
, which we already used to simplify the expression of $F(r)$. If we
analytically continue the coordinate $t\rightarrow i\chi $ and then perform
further analytic continuations in the $H^{2}$ sector, the following distinct
metrics are obtained: 
\begin{eqnarray}
ds^{2} &=&F(r)(d\chi +l\cos td\phi
)^{2}+F^{-1}(r)dr^{2}+(r^{2}+l^{2}/4)(-dt^{2}+\sin ^{2}td\phi
^{2})+r^{2}dy^{2}  \nonumber \\
ds^{2} &=&F(r)(d\chi +l\sinh \theta
dt)^{2}+F^{-1}(r)dr^{2}+(r^{2}+l^{2}/4)(d\theta ^{2}-\cosh ^{2}\theta
dt^{2})+r^{2}dy^{2},  \nonumber \\
ds^{2} &=&F(r)(d\chi +l\cosh \theta
dt)^{2}+F^{-1}(r)dr^{2}+(r^{2}+l^{2}/4)(d\theta ^{2}-\sinh ^{2}\theta
dt^{2})+r^{2}dy^{2},  \nonumber \\
ds^{2} &=&F(r)(d\chi +le^{\theta
}dt)^{2}+F^{-1}(r)dr^{2}+(r^{2}+l^{2}/4)(d\theta ^{2}-e^{2\theta
}dt^{2})+r^{2}dy^{2}.  \label{adsbubbles5d}
\end{eqnarray}%
They are solutions of the vacuum Einstein field equations with negative
cosmological constant $\Lambda =-6/l^{2}$.

For the last three geometries the coordinate $\theta $ is no longer periodic
and can take any real value. The geometry in the second bracket is described
by a two-dimensional AdS space. As is well known, this space can have
non-trivial identifications and so the $2$-dimensional sector can describe a
$2$-dimensional black hole (as in the second and third metrics above), while the first metric describes pure AdS in standard coordinates. Notice
however that in this case, the geometry of a fixed $(\chi ,r,y)$-slice is $%
AdS_{2}$ modified by the term $F(r)l^{2}\sinh ^{2}\theta dt^{2}$ as an
effect of the non-trivial fibration over the $(\theta ,t)$-sector. This
extra term will vanish only at points where $F(r)=0$ (thence, on the bubble).

After a coordinate transformation the last three metrics can be written in
a compact form 
\begin{eqnarray}
ds^{2}&=&F(r)(d\chi +lxdt)^{2}+F^{-1}(r)dr^{2}+(r^{2}+l^{2}/4)\left( \frac{%
dx^{2}}{x^{2}+k}-(x^{2}+k)dt^{2}\right)  \nonumber \\
&&+r^{2}dy^{2},  \label{bh5d1}
\end{eqnarray}
with $k=-1,1,0$. They are all locally equivalent under changes of
coordinates. However, depending on the identifications made, the global
structure can be quite different.

The quartic function in the numerator of $F(r)$ can have only two real roots
--- for $m>0$, one is positive (denoted by $r_{+}$) and the other one is
negative (denoted by $r_{-}$): 
\begin{eqnarray}
r_{\pm }=\pm \frac{l}{2}\sqrt{\sqrt{1+64m/l^{2}}-1}.
\end{eqnarray}
The conical singularities at either root of $F(r)$ in the $(\chi ,r)$-sector
can be eliminated if the periodicity of the $\chi $-coordinate is 
\begin{eqnarray}
\beta =\frac{4\pi }{|F^{\prime }(r_{\pm })|}=\frac{2\pi l}{\sqrt{\sqrt{%
1+64m/l^{2}}-1}}.  \label{betachi}
\end{eqnarray}

\begin{figure}[htbp]
\centering 
\begin{minipage}[c]{.45\textwidth}
         \centering
         \includegraphics[width=\textwidth,angle=0,keepaspectratio]{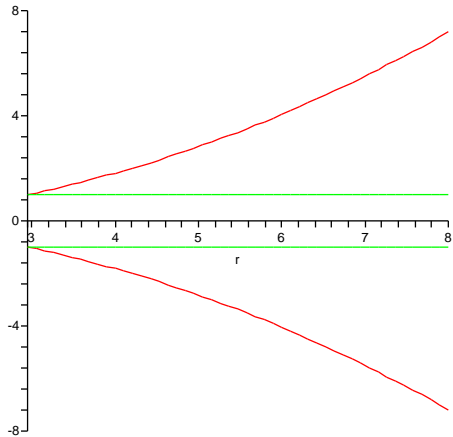}
\end{minipage}
\caption{Ergoregion of the topological metric with $k=-1$ in the $(r,x)$
plane for $m=20$ and $l=1$. The ergoregion is confined to the infinite
regions bounded by the red curves and outside the black hole horizons $|x|=1$%
.}
\label{ergo}
\end{figure}

Now, for $r>r_{+}$ (or $r<r_{-}$) the first three metrics will describe
stationary backgrounds. Note that the metric (\ref{bh5d1}) is stationary and
it possesses the Killing vector $\xi =\frac{\partial }{\partial t}$. The
norm of this Killing vector is 
\[
\xi \cdot \xi =l^{2}x^{2}F(r)-(r^{2}+l^{2}/4)(x^{2}+k),
\]%
and we find that in general there is an ergoregion iff 
\[
\left( \frac{4l^{2}F(r)}{4r^{2}+l^{2}}-1\right) x^{2}>k.
\]%
However, since the expression in the bracket is always negative we find that
there exists an ergoregion only if $k=-1$ in which case the following
constraint is obtained: 
\[
|x|<\frac{4r^{2}+l^{2}}{l\sqrt{64m+l^{2}}}.
\]%
The ergoregion corresponds to a strip in the $(r,x)$ plane bounded by the
horizons located at $|x|=1$ and two curves that asymptote to $1$ for $%
r\rightarrow r_{\pm }$, while for large values of $r$ the strip will largely
broaden (see figure \ref{ergo}) and the curves asymptote to $\frac{4x^{2}}{l%
\sqrt{64m+l^{2}}}$. Also, as it is apparent from the above formula, the
strip broadens as $m$ decreases.

In the remaining cases, for $k=0,1$ there is no ergoregion.

The asymptotic structure of the above metrics is 
\begin{eqnarray}
ds^2&=&r^2/l^2(d\chi+lxdt)^2+l^2/r^2dr^2+r^2\left(\frac{dx^2}{x^2+k}%
-(x^2+k)dt^2\right)+r^2dy^2.
\end{eqnarray}
Now, it is easy to read the boundary geometry --- up to a conformal
rescaling factor $r^2/l^2$, the boundary metric is 
\begin{eqnarray}
ds^2&=&l^2(d\tilde{\chi}+xdt)^2+l^2\left(\frac{dx^2}{x^2+k}%
-(x^2+k)dt^2\right)+l^2dy^2.
\end{eqnarray}
Here, we use a rescaled coordinate $\tilde{\chi}=\chi/l$. From (\ref{betachi}%
) it is easy to see that if $m=\frac{l^2s^2(s^2+8)}{1024}$ then $\tilde{\chi}
$ has periodicity $4\pi/s$, with $s$ an integer. Remarkably, for $s=1$ the
boundary geometry is conformally flat. This can be easily seen from the fact
that the boundary metric is the product of a $3$-dimensional space of
constant curvature (i.e. pure $AdS_3$) with a line (or a circle if we also
compactify the $y$ coordinate). Furthermore, one can also make nontrivial
identifications in the $AdS_3$ sector which turn it into the $BTZ$ black
hole. We will have to say more about these solutions in the Discussion
section.

Finally, let us consider the first metric from (\ref{adsbubbles5d}). Even if
formally it can be transformed into the $k=-1$ metric by a coordinate
transformation\footnote{Notice however that the range of the $x$ coordinate is different: for the $%
k=-1$ metric $x\geq 1$, while for the first metric in (\ref{adsbubbles5d})
we must take $|x|\leq 1$. Another difference is that the periodicity in $\chi $ would be related to a periodicity in $\phi$ for the first metric, while for the second one, with $k=-1$, it would require us to make $t$ periodic. Therefore we find that while there are no hyperbolic Misner
strings for $k=-1$, the fourth metric could have Misner strings.}, if $\phi$
is periodic then the global structure of these spaces is completely
different. The geometry in the $(\chi,t,\phi)$-sector resembles the usual Hopf-type fibration. The $\chi $-circle is now fibred over the circle described by $\phi $. However, the fibration is twisted as a
function of time. At $t=0$ we have a pair of orthogonal circles provided we define $\chi $ appropriately. As time increases we have the $\chi $-circle twisting around relative to the $\phi$-circle, while the $\phi $-circle is getting bigger. The latter reaches a maximum, and then begins to shrink. However the $\chi $-circle is still twisting, and by the time the $\phi $-circle has shrunk back to zero, the $\chi $-circle has twisted only `halfway' round.  Over this cycle the integral $\int d(l\cos t d\phi)$ is well-defined,
and it equals $4\pi \ell $ since we are integrating $t$ from $0$ to $\pi $. This will set the periodicity of $\chi$ to be $4\pi l/s$, where $s$ is an integer. Recall now that the quartic function $F(r)$ can have only two real roots, one positive ($r_+$) and one negative ($r_-$) for $m>0$. If $\phi$ is an angular coordinate with period $2\pi$, then in order to eliminate the Misner string singularity we require that the period $\beta=4\pi/|F^{\prime}(r_{\pm})|$ be equal to $4\pi l/s$, where $s$ is some integer. This
further restricts the value of the mass parameter such that $m=\frac{l^2s^2(s^2+8)}{1024}$.

For $r>r_{+}$ (or $r<r_{-}$), this metric describes then a bubble located at 
$r=r_+$, which expands from zero size to a finite size and then contracts to
zero size again. All the spacetime events are causally connected with each
other. Near the initial expansion (or the final contraction) the scale
factor is linear in time and the spacetime expands or contracts like a Milne
universe. The boundary geometry for this bubble spacetime is given by 
\begin{eqnarray}
ds^2&=&(d\chi+l\cos td\phi)^2+l^2(-dt^2+\sin^2t d\phi^2)+l^2dy^2,
\end{eqnarray}
where $\chi$ is periodic with period $4\pi l/s$.

\subsection{Nutty bubbles in dS}

The Taub-NUT ansatz that we shall use in the construction of these spaces is
the following: 
\begin{equation}
ds^{2}=-F(r)(dt-2n\cos \theta d\varphi
)^{2}+F^{-1}(r)dr^{2}+(r^{2}+n^{2})(d\theta ^{2}+\sin ^{2}\theta d\varphi
^{2})+r^{2}dz^{2}.  \label{TNB5d}
\end{equation}%
The above metric will be a solution of the Einstein field equations with
positive cosmological constant $\Lambda =\frac{6}{l^{2}}$ provided 
\begin{equation}
F(r)=\frac{4ml^{2}-r^{4}-2n^{2}r^{2}}{l^{2}(r^{2}+n^{2})},  \label{5dSF}
\end{equation}%
where the field equations impose the constraint\footnote{%
In the following we shall use this constraint to eliminate $n$ from the
metric.} $4n^{2}=l^{2}$. Notice that, for large values of $r$, the function $%
F(r)$ takes negative values and $r$ becomes effectively a timelike
coordinate as one should expect in a region outside the cosmological
horizon. In order to remove the usual Misner string singularity in the
metric, we have to assume that the coordinate $t$ is periodic with period $%
4\pi l$. If we analytically continue the coordinate $t\rightarrow i\chi $
and one of the coordinates in the $S^{2}$ sector we obtain the following
metrics: 
\begin{eqnarray}
ds^{2} &=&F(r)(d\chi +l\cos \theta
dt)^{2}+F^{-1}(r)dr^{2}+(r^{2}+l^{2}/4)(d\theta ^{2}-\sin ^{2}\theta
dt^{2})+r^{2}dy^{2},  \nonumber \\
ds^{2} &=&F(r)(d\chi +l\cosh td\phi
)^{2}+F^{-1}(r)dr^{2}+(r^{2}+l^{2}/4)(-dt^{2}+\sinh ^{2}td\phi
^{2})+r^{2}dy^{2},  \nonumber \\
ds^{2} &=&F(r)(d\chi +l\sinh td\phi
)^{2}+F^{-1}(r)dr^{2}+(r^{2}+l^{2}/4)(-dt^{2}+\cosh ^{2}td\phi
^{2})+r^{2}dy^{2},  \nonumber \\
ds^{2} &=&F(r)(d\chi +le^{t}d\phi
)^{2}+F^{-1}(r)dr^{2}+(r^{2}+l^{2}/4)(-dt^{2}+e^{2t}d\phi ^{2})+r^{2}dy^{2}.
\label{5dsbubbles}
\end{eqnarray}%
These metrics satisfy the vacuum Einstein field equations with positive
cosmological constant $\Lambda =6/l^{2}$, where $F(r)$ is given by (\ref%
{5dSF}). However, for large values of $r$ the function $F(r)$ becomes
negative and the signature of the spacetime will change accordingly. To
avoid this situation one possibility is to consider two roots of the
function $F(r)$ and to restrict the values of the $r$ coordinate such that $%
F(r)$ is always positive. Namely we restrict the range of the $r$ coordinate
such that $r_{-}<r<r_{+}$, where $r_{\pm }$ are two roots of $F(r)$ and in
this way we avoid the change in the metric signature. It is easy to see that
if $m>0$ then $F(r)$ has two real roots only if 
\[
r_{\pm }=\pm \frac{l}{2}\sqrt{\sqrt{1+64m/l^{2}}-1}\,.
\]%
Fortunately, the conical singularities at the roots of $F(r)$ in the $(\chi
,r)$-sector can both be eliminated in the same time if we choose the
periodicity of the $\chi $-coordinate to be given by 
\[
\beta =\frac{4\pi }{|F^{\prime }(r_{\pm })|}=\frac{2\pi l}{\sqrt{\sqrt{%
1+64m/l^{2}}-1}}\,.
\]%
To eliminate the Misner string singularity in the first metric in (\ref%
{5dsbubbles}), we require that the period $\beta $ be equal to $4\pi l/s$,
where $s$ is some integer. This further restricts the value of the mass
parameter such that $m=\frac{l^{2}s^{2}(s^{2}+8)}{1024}$. Notice that,
locally, all these metrics are equivalent, being related by coordinate
transformations. However, these spaces will be equivalent globally only if
the coordinate $\phi $ is unwrapped. At every fixed $(\chi ,r,y)$ the
geometry is that of a perturbed two-dimensional de Sitter spacetime as an
effect of the non-trivial fibration.

To better understand the geometry in the $(\chi ,r,y)$ sector let us focus
on a section with $t,\phi $ held fixed. Then the metric in the $(\chi ,r,y)$%
-sector becomes 
\[
ds^{2}=F(r)d\chi ^{2}+F^{-1}(r)dr^{2}+r^{2}dy^{2}. 
\]%
We restrict the values of the $r$-coordinate between the two roots of $F(r)$
and since they have the same magnitude, we shall take $%
r_{+}^{2}=r_{-}^{2}=r_{0}^{2}$. Then, it is easy to see that we can write 
\[
F(r)=(r_{0}^{2}-r^{2})\frac{4r^{2}+2l^{2}+4r_{0}^{2}}{l^{2}(4r^{2}+l^{2})}%
=(r_{0}^{2}-r^{2})f(r), 
\]%
where $f(r)$ is strictly positive everywhere. Now if we make the following
change of coordinates 
\[
r^{2}=r_{0}^{2}(1-x^{2}), 
\]%
the metric in the $(\chi ,r,y)$ sector becomes 
\[
ds^{2}=\frac{dx^{2}}{(1-x^{2})f(r_{0}\sqrt{1-x^{2}})}%
+r_{0}^{2}(1-x^{2})dy^{2}+r_{0}^{2}x^{2}f(r_{0}\sqrt{1-x^{2}})d\chi ^{2}. 
\]
A further change of coordinates $x=\sin \psi $ will bring it in the form: 
\[
ds^{2}=\frac{d\psi ^{2}}{f(r_{0}\cos \psi )}+r_{0}^{2}\cos ^{2}\psi
dy^{2}+r_{0}^{2}\sin ^{2}\psi f(r_{0}\cos \psi )d\chi ^{2}, 
\]%
where 
\[
f(r_{0}\cos \psi )=\frac{1}{l^{2}}\bigg[1+\frac{4r_{0}^{2}}{4r_{0}^{2}\cos
^{2}\psi +l^{2}}\bigg]. 
\]%
It can be easily seen that the geometry in this sector is one of a deformed $%
3$-sphere. We conclude that our bubble metrics describe non-trivial fibrations
of a $3$-sphere over a $2$-dimensional dS space.

If the coordinate $\phi$ is periodic, the circle geometry that it describes
will evolve differently for each of the above geometries. For instance, for
the second metric in (\ref{5dsbubbles}) the evolution is that of a circle
that begins with zero radius at $t=0$ and then expands exponentially as $%
t\rightarrow \infty$, while for the third metric we obtain a de Sitter
evolution of a circle with exponentially large radius at $t\rightarrow
-\infty$ that exponentially shrinks to a minimal value and then expands
again. Finally the fourth geometry describes the evolution of a circle which
begins with zero radius at $t\rightarrow -\infty$ and then expands
exponentially as $t\rightarrow \infty$. Similar with the four-dimensional
situation considered in \cite{nuttybubbles}, a null curve in a geometry for
which the bubbles are expanding has $|\dot{\phi}|\leq e^{-t}|\dot{t}|$ at
late times, where the overdot refers to a derivative with respect to proper
time. Hence, observers at different values of $\phi$ will eventually lose
causal contact. On the other hand null rays at fixed $\phi$ and $y$ obey the
relation 
\begin{eqnarray}
\dot{r}^2+V(r)=0,
\end{eqnarray}
where $V(r)=p_{\chi}^2+4p_y^2F(r)/r^2-4E^2F(r)/(4r^2+l^2)$ is an effective
potential, $p_{\chi}=\dot{\chi}F$ is the conserved momentum along the $\chi$
direction, $p_y=r^2\dot{y}$ is the conserved momentum along the $y$
direction and $E=(r^2+l^2/4)\dot{t}$ is the conserved energy. Generically,
if $p_y=0$ then the null geodesics oscillate between some minimal and
maximal values of $r$, which can be chosen to be within the admitted range
of $r$. Hence observers at any two differing values of $r$ can be causally
connected. However, if the observers are at different values of $y$ then the
effective potential diverges at $r=0$, which means that there will be two
regions that can be causally disconnected. It is easy to check that it is
possible for observers at different values of $\chi$, respectively $y$ to be
causally connected at fixed values of $r$.

Let us notice that the first metric from (\ref{5dsbubbles}) is stationary
since it possesses the Killing vector $\xi=\frac{\partial}{\partial t}$. The
norm of this Killing vector is 
\begin{eqnarray}
\xi\cdot\xi&=&l^2F(r)-(r^2+l^2/4-l^2F(r))\sin^2\theta
\end{eqnarray}
and so it becomes spacelike unless $\hat{\theta}(r)\leq |\theta|\leq \pi-%
\hat{\theta}(r)$. Here, we use the notation 
\begin{eqnarray}
\hat{\theta}(r)&=&\tan^{-1}\left(\frac{2l\sqrt{F(r)}}{\sqrt{4r^2+l^2}}%
\right).
\end{eqnarray}
As in the four-dimensional case \cite{nuttybubbles} these limits will
describe an `ergocone' for the space-time. The above angle vanishes at $%
r=r_{\pm}$ while it attains a maximum value at $r=0$.

\subsection{Nutty Rindler bubbles in flat backgrounds}

We can also obtain NUT spaces with non-trivial topology if we construct the
circle fibration over a two-dimensional torus $T^2$, 
\begin{equation}
ds^{2}=-F(r)(dt-2n\theta d\phi )^{2}+F^{-1}(r)dr^{2}+(r^{2}+n^{2})(d\theta
^{2}+d\phi ^{2})+r^{2}dy^{2}  \label{54}
\end{equation}%
where 
\begin{equation}
F(r)=\frac{4ml^{2}+r^{4}+2n^{2}r^{2}}{l^{2}(r^{2}+n^{2})}  \label{55}
\end{equation}%
and the constraint equation takes now the form $\Lambda n^2=0$. Consistent
Taub-NUT spaces with toroidal topology exist if and only if the cosmological
constant vanishes. The Euclidean version of this solution, obtained by
analytic continuation of the coordinate $t\rightarrow it$ and of the
parameter $n\rightarrow in$, has a curvature singularity at $r=n$. Note that
if we consider $n=0$ in the above constraint we obtain the AdS/dS black hole
solution in five dimensions with toroidal topology.

If the cosmological constant vanishes, then we can have $n\neq 0$ and the
metric becomes 
\begin{equation}
ds^{2}=-F(r)(dt-2n\theta d\phi )^{2}+F^{-1}(r)dr^{2}+(r^{2}+n^{2})(d\theta
^{2}+d\phi ^{2})+r^{2}dy^{2},  \label{55a}
\end{equation}%
where 
\begin{equation}
F(r)=\frac{4m}{r^{2}+n^{2}}.
\end{equation}%
The asymptotic structure of the above metric is given by 
\begin{equation}
ds^{2}=\frac{4m}{r^{2}}(dt-2n\theta d\phi )^{2}+\frac{r^{2}}{4m}%
dr^{2}+r^{2}(d\theta ^{2}+d\phi ^{2}+dy^{2}).
\end{equation}%
If $y$ is an angular coordinate then the angular part of the metric
parameterizes a three torus. The Euclidian section of the solution described
by (\ref{55a}) is not asymptotically flat and has a curvature singularity
localized at $r=0$. However, let us notice that for $r\leq n$ the signature
of the space becomes completely unphysical. Hence, for the Euclidian
section, we should restrict the values of the radial coordinate such that $%
r\geq n$.

Consider now the analytic continuations of the coordinates $t\rightarrow
i\chi $ and $\theta \rightarrow -it$ (respectively $\phi \rightarrow -it$)
in the case of fibration over a torus. We obtain the spacetimes%
\begin{eqnarray}
ds^{2} &=&F(r)(d\chi +2ntd\phi
)^{2}+F^{-1}(r)dr^{2}+(r^{2}+n^{2})(-dt^{2}+d\phi ^{2})+r^{2}dy^{2}, 
\nonumber \\
ds^{2} &=&F(r)(d\chi +2n\theta dt)^{2}+F^{-1}(r)dr^{2}+(r^{2}+n^{2})(d\theta
^{2}-dt^{2})+r^{2}dy^{2}
\end{eqnarray}%
whose metrics are locally equivalent under coordinate transformations.
However, if one of the coordinates $\phi$ or $\theta$ is periodic then they
represent globally different spaces. Another metric --- related to the above
by coordinate transformations --- is a generalization to five-dimensions of
the four dimensional nutty Rindler spacetime. Since $F(r)$ has no roots
these spaces are not really bubbles. However they still represent
interesting time-dependent backgrounds, with metrics given by 
\begin{eqnarray}
ds^{2} &=&F(r)(d\chi +nt^{2}d\phi
)^{2}+F^{-1}(r)dr^{2}+(r^{2}+n^{2})(-dt^{2}+t^{2}d\phi ^{2})+r^{2}dy^{2}, 
\nonumber \\
ds^{2} &=&F(r)(d\chi +n\theta
^{2}dt)^{2}+F^{-1}(r)dr^{2}+(r^{2}+n^{2})(d\theta ^{2}-\theta
^{2}dt^{2})+r^{2}dy^{2}.
\end{eqnarray}%
Let us notice that, by analogy with the four-dimensional case, in the first
case, the geometry of a slice $(r,\theta ,y)$ is that of a twisted torus
which has a Milne-type evolution. The second geometry describes a static
spacetime (with the Killing vector $\xi =\frac{\partial }{\partial t}$) with
an ergoregion described by 
\[
\theta ^{2}>\frac{r^{2}+n^{2}}{n^{2}F(r)}. 
\]
For large values of $r$, the ergoregion includes almost the entire $(\theta
,r)$ plane except for a strip bounded by two curves, opposite the $r$-axis,
which asymptote to parabolas. For small values of $r$ the strip narrows and
the boundary curves asymptote to $\pm n/\sqrt{2m}$. For the first geometry,
as $t^2\dot{\phi}^2\leq \dot{t}^2$ we have $\phi\sim \ln t$ and observers
with different values of $\phi $ can communicate with each other for
arbitrarily large $t$. In the second geometry we obtain $\dot{\theta}%
^2\leq\theta^2\dot{t}^2$, \textit{i.e.} $\theta\sim e^{|t|}$ and we see that
there is no restriction as to the maximum change of coordinate $\theta$ for
points on the null curve as $t\rightarrow\pm\infty$ and observers at points
with different $\theta $ can communicate with each other.

\section{Higher dimensional Nutty Bubbles}

\label{higherdim}

We now consider some of the higher dimensional Taub-NUT spaces constructed
recently in \cite{csrm,Page1,TNletter}. As mentioned previously, in the
Taub-NUT ansatz the idea is to construct such spaces as radial extensions of 
$U(1)$-fibrations over an even dimensional base $B$ endowed with an
Einstein-K\"ahler metric $g_{B}$. In a $(2k+2)$-dimensional Taub-NUT
space the base factor over which one constructs the circle fibration can
have at most dimension $2k$, in which case the metric is 
\begin{equation}
F^{-1}(r)dr^{2}+(r^{2}+N^{2})g_{B}-F(r)(dt+2N\mathcal{A})^{2},  \label{I1}
\end{equation}%
where $t$ is the coordinate on the fibre and $\mathcal{F}=d\mathcal{A}$ is
the K\"ahler $2$-form. Here $N$ is the NUT charge and $F(r)$ is a
function of $r$. More generally, in the even-dimensional cases we can
consider circle fibrations over base spaces that can be factorized in the
form $B=M_{1}\times \dots \times M_{k}$ where $M_{i}$ are Einstein-K$\ddot{a}
$hler manifolds. In this case we can associate a NUT charge $N_{i}$ for
every such base factor $M_i$. Also, in the above ansatz we replace $%
(r^{2}+N^{2})g_{B}$ with the sum $ \sum_{i}(r^{2}+N_{i}^{2})g_{M_{i}}$ and $%
2N\mathcal{A}$ by $\sum_i2N_i\mathcal{A}_i$. In particular, we can use the
sphere $S^{2}$, the torus $T^{2}$ or the hyperboloid $H^{2}$ or in general $%
CP^n$ as factor spaces.

More generally one can consider more general Taub-NUT-like spaces with
factorizations of the base space of the form $B=\prod_i M_i\times Y$, where
each factor $M_i$ is endowed with an Einstein-K\"ahler metric $g_{M_i}
$ while $Y$ is a general Einstein space with metric $g_{Y}$. In these cases
one can consider the $U(1)$-fibration only over the factored space $%
M=\prod_iM_i$ of the base $B$ and take then a warped product with the
manifold $Y$. Quite generically, we can associate a nut parameter $N_i$ with
every such factor $M_i$ and the general ansatz is then given by 
\begin{equation}
F^{-1}(r)dr^{2}+\sum_{i}(r^{2}+N_{i}^{2})g_{M_{i}}+r^{2}g_{Y}-F(r)(dt+%
\sum_i2N_i\mathcal{A}_i)^{2}.  \label{l2}
\end{equation}
We now consider particular cases of these ans\"atze. To be more specific we
shall focus on a couple of six-dimensional metrics.

\subsection{Bubbles in flat backgrounds}

In six dimensions the base space is four-dimensional and we can use products
of the form $M_{1}\times M_{2}$ of two-dimensional Einstein-K\"ahler
spaces or we can use $CP^{2}$ as a four-dimensional base space over which to
construct the circle fibrations. If we use products of two dimensional
Einstein-K\"ahler spaces then we can consider all the cases in which $%
M_{i}$, $i=1,2$ can be a sphere $S^{2}$, a torus $T^{2}$ or a hyperboloid $%
H^{2}$. The circle fibration can be constructed over the whole base space $%
M_{1}\times M_{2}$, in which case we can have two distinct nut parameters
associated with each factor $M_i$ or, in the case of metrics with only one
nut parameter, just over one factor space $M_{1}$, in which case we also
take the warped product with $M_2$ as in (\ref{l2}).

We shall consider first the case in which $M_{1}=M_{2}=S^{2}$ and assume
that the $U(1)$ fibration is constructed over the whole base space $%
S^{2}\times S^{2}$. Then the corresponding six-dimensional Taub-NUT solution
is given by \cite{csrm} 
\begin{eqnarray}
ds^{2} &=&-F(r)(dt-2n_{1}\cos \theta _{1}d\varphi _{1}-2n_{2}\cos \theta
_{2}d\varphi _{2})^{2}+F^{-1}(r)dr^{2}  \nonumber \\
&&+(r^{2}+n_{1}^{2})(d\theta _{1}^{2}+\sin ^{2}\theta _{1}d\varphi
_{1}^{2})+(r^{2}+n_{2}^{2})(d\theta _{2}^{2}+\sin ^{2}\theta _{2}d\varphi
_{2}^{2}),
\end{eqnarray}%
where 
\begin{eqnarray}
F(r) &=&\frac{%
3r^{6}+(l^{2}+5n_{2}^{2}+10n_{1}^{2})r^{4}+3(n_{2}^{2}l^{2}+10n_{1}^{2}n_{2}^{2}+n_{1}^{2}l^{2}+5n_{1}^{4})r^{2}%
}{3(r^{2}+n_{1}^{2})(r^{2}+n_{2}^{2})l^{2}}  \nonumber \\
&&-\frac{6ml^{2}r+3n_{1}^{2}n_{2}^{2}(l^{2}+5n_{1}^{2})}{%
3(r^{2}+n_{1}^{2})(r^{2}+n_{2}^{2})l^{2}}.  \label{6nn}
\end{eqnarray}%
Here the above metric is a solution of vacuum Einstein field equations with
cosmological constant ($\lambda =-\frac{10}{l^{2}}$) if and only if $%
(n_{1}^{2}-n_{2}^{2})\lambda =0$. Consequently, we see that differing values
for $n_{1}$ and $n_{2}$ are possible only if the cosmological constant
vanishes. For $n_{1}=n_{2}=n$ the above solution reduces to the
six-dimensional solution found and studied in \cite%
{Bais,Page,Awad,Akbar,Robinson,Lorenzo}. In the case of only one nut charge $%
n$ there are no consistent analytic continuations of the coordinates that
lead to acceptable time-dependent metrics with Lorentzian signature \cite%
{nuttybubbles}. Basically, the reason for this is that if we perform
analytic continuations of the coordinates on one factor space $S^{2}$ we
also have to send $n\rightarrow in$, which will force us to analytically
continue the coordinates in the second sphere $S^{2}$ yielding spaces with
two timelike directions. However, if the nut parameters are independent then
we can analytically continue the coordinates in one factor $M_{i}$ only and
analytically continue the nut parameter associated with the second factor $%
M_{j}$. This enables us to construct nutty bubble spacetimes in virtually
any dimension. For this reason, in what follows we shall look at the case of
two different nut charges, that is we set the cosmological constant to zero.
Let us consider the Euclidean section, obtained by the following analytic
continuations $t\rightarrow i\chi $ and $n_{j}\rightarrow in_{j}$ where $%
j=1,2$: 
\begin{eqnarray}
ds^{2} &=&F_{E}(r)(d\chi -2n_{1}\cos \theta _{1}d\varphi _{1}-2n_{2}\cos
\theta _{2}d\varphi _{2})^{2}+F_{E}^{-1}(r)dr^{2}  \nonumber \\
&&+(r^{2}-n_{1}^{2})(d\theta _{1}^{2}+\sin ^{2}\theta _{1}d\varphi
_{1}^{2})+(r^{2}-n_{2}^{2})(d\theta _{2}^{2}+\sin ^{2}\theta _{2}d\varphi
_{2}^{2}),  \nonumber \\
F_{E}(r) &=&\frac{r^{4}-3(n_{1}^{2}+n_{2}^{2})r^{2}-6mr-3n_{1}^{2}n_{2}^{2}}{%
3(r^{2}-n_{1}^{2})(r^{2}-n_{2}^{2})}.
\end{eqnarray}%
This metric is a solution of the vacuum Einstein field equations without
cosmological constant, for any values of the parameters $n_{1}$ and $n_{2}$.
We set $n_{1}>n_{2}$ without loss of generality. In this case in the
Euclidean section the radius $r$ cannot be smaller than $n_{1}$ or the
signature of the spacetime will change. The Taub-nut solution in
this case corresponds to a two-dimensional fixed-point set located at $%
r=n_{1}$. There is still a curvature singularity located at $r=n_{1}
$. \ While superficially it would seem that this could removed by
setting the periodicity of the coordinate $\chi $ to be $8\pi
n_{1}$ (thereby setting $m=m_{p}=-\frac{n_{1}^{3}+3n_{1}n_{2}^{2}}{%
3}$), a more careful analysis reveals that this nut solution is
actually still singular. This is because the nut parameters $n_{1,2}$
must be rationally related, in which case the periodicity of the $%
\chi $ coordinate is $8\pi n_{2}/k$, where $k$
is an integer. As $n_{2}<n_{1}$ it is not possible to match this
periodicity with $8\pi n_{1}$ for any integer $k$.

On the other hand, the bolt solution corresponds to $r\geq r_0>n_1$ and the
periodicity is found to be $\frac{4\pi}{|F_E^{\prime}(n_1)|}=4\pi r_0$. It
is now possible to match it with $8\pi n_2/k$ with $k$ the integer and we
obtain $r_0=\frac{2n_2}{k}$. The bolt solution is then non-singular as long
as $r_0>n_1$, that is for $k=1$ and $n_1<2n_2$.

We are now ready to perform analytic continuations on the sphere factors in
order to generate new time-dependent backgrounds. For instance, we can
consider $\theta _{1}\rightarrow it+\frac{\pi }{2}$, which will force us to
take $n_{1}\rightarrow in_{1}$. We obtain the following time dependent
solution: 
\begin{eqnarray}
ds^{2} &=&\tilde{F}_{E}(r)(d\chi +2n_{1}\sinh td\phi _{1}+2n_{2}\cos \theta
_{2}d\phi _{2})^{2}+\tilde{F}_{E}^{-1}(r)dr^{2}  \nonumber \\
&&+(r^{2}+n_{1}^{2})(-dt^{2}+\cosh ^{2}td\phi
_{1}^{2})+(r^{2}-n_{2}^{2})(d\theta _{2}^{2}+\sin ^{2}\theta _{2}d\phi
_{2}^{2}),  \nonumber \\
\tilde{F}_{E} &=&\frac{%
r^{4}+3(n_{1}^{2}-n_{2}^{2})r^{2}-6rm+3n_{1}^{2}n_{2}^{2}}{%
3(r^{2}+n_{1}^{2})(r^{2}-n_{2}^{2})}.  \label{tildes2}
\end{eqnarray}%
More generally, as with the four-dimensional case,  after performing
appropriate analytic continuations we end up with the following metrics: 
\begin{eqnarray}
ds^{2} &=&\tilde{F}_{E}(r)(d\chi +2n_{1}\cosh td\phi _{1}+2n_{2}\cos \theta
_{2}d\phi _{2})^{2}+\tilde{F}_{E}^{-1}(r)dr^{2},  \nonumber \\
&&+(r^{2}+n_{1}^{2})(-dt^{2}+\sinh ^{2}td\phi
_{1}^{2})+(r^{2}-n_{2}^{2})(d\theta _{2}^{2}+\sin ^{2}\theta _{2}d\phi
_{2}^{2}),  \nonumber \\
ds^{2} &=&\tilde{F}_{E}(r)(d\chi +2n_{1}e^{t}d\phi _{1}+2n_{2}\cos \theta
_{2}d\phi _{2})^{2}+\tilde{F}_{E}^{-1}(r)dr^{2},  \nonumber \\
&&+(r^{2}+n_{1}^{2})(-dt^{2}+e^{2t}d\phi _{1}^{2})+(r^{2}-n_{2}^{2})(d\theta
_{2}^{2}+\sin ^{2}\theta _{2}d\phi _{2}^{2}),  \nonumber \\
ds^{2} &=&\tilde{F}_{E}(r)(d\chi +2n_{1}\cos \theta dt+2n_{2}\cos \theta
_{2}d\phi _{2})^{2}+\tilde{F}_{E}^{-1}(r)dr^{2},  \nonumber \\
&&+(r^{2}+n_{1}^{2})(d\theta _{1}^{2}-\sin ^{2}\theta
_{1}dt^{2})+(r^{2}-n_{2}^{2})(d\theta _{2}^{2}+\sin ^{2}\theta _{2}d\phi
_{2}^{2}).  \label{6dS2}
\end{eqnarray}%
They are also solutions of vacuum Einstein field equations with the same
function $\tilde{F}_{E}$ as in (\ref{tildes2}).

While locally all these spaces are equivalent under coordinate
transformations, if we compactify the coordinate $\phi _{1}$(respectively $%
\theta _{1}$ for the last metric in (\ref{6dS2})) the global structure and
in particular the evolution of the bubble will be different. The bubble will
be located at the biggest root $r_{0}$ of $\tilde{F}_{E}(r)$ such that $%
r_{0}>n_{2}$ and we also restrict the range of the $r$ coordinate such that $%
r\geq n_{2}$. Elimination of the Misner string sets the periodicity
of the $\chi $ coordinate to be $8\pi n_{2}/k$, which
in turn must be matched with the periodicity $4\pi /|\tilde{F}_{E}^{\prime
}(r_{0})|$, introduced after we eliminate any possible conical
singularities in the $(\chi ,r)$ sector. Again we have two solutions: a nut
and a bolt.

The nut solution corresponds to $r_{0}=n_{2}$ and in this case the mass
parameter is $n_{2}(3n_{1}^{2}-n_{2}^{2})/3$. Notice that the mass parameter
can have either sign. The coordinate $\chi $ has periodicity $8\pi n_{2}$.
It is very interesting to note that the fixed-point set of the isometry
generated by $\partial /\partial \chi $ is effectively two dimensional. The
induced geometry on the `bubblenut' is a two-dimensional de Sitter
space, whose metrics are one of the following 
\begin{eqnarray}
ds_{2}^{2} &=&(n_{1}^{2}+n_{2}^{2})(-dt^{2}+\cosh ^{2}td\phi _{1}^{2}), 
\nonumber \\
ds_{2}^{2} &=&(n_{1}^{2}+n_{2}^{2})(-dt^{2}+\sinh ^{2}td\phi _{1}^{2}), 
\nonumber \\
ds_{2}^{2} &=&(n_{1}^{2}+n_{2}^{2})(-dt^{2}+e^{2t}d\phi _{1}^{2}),  \nonumber
\\
ds_{2}^{2} &=&(n_{1}^{2}+n_{2}^{2})(d\theta _{1}^{2}-\sin ^{2}\theta
_{1}dt^{2}).
\end{eqnarray}%
which differ globally but not locally. If the coordinate $\phi _{1}$
is periodically identified then at any fixed time $r=n_{2}$ is our
`bubblenut': a circle with minimal circumference that expands or contracts.
The first three de Sitter geometries above correspond to three different
evolutions of this circle: the first geometry describes the evolution of a
circle with exponentially large radius at $t\rightarrow -\infty $, which
shrinks to a minimal value and expands exponentially again for $t\rightarrow
\infty $; the second geometry describes the evolution of a circle which
begins with zero radius at $t=0$ and expands exponentially, while the third
geometry describes a circle that begins with exponentially small radius at $%
t\rightarrow -\infty $ and then expands exponentially. The last geometry is
stationary as in these coordinates the metric has a Killing vector $\xi
=\partial /\partial t$. The norm of this Killing vector is 
\[
\xi \cdot \xi =4n_{1}^{2}\tilde{F}_{E}(r)-(r^{2}+n_{1}^{2}-4n_{1}^{2}\tilde{F%
}_{E}(r))\sin ^{2}\theta _{1}
\]%
so that it will become spacelike unless $\hat{\theta _{1}}(r)\leq |\theta
_{1}|\leq \pi -\hat{\theta _{1}}(r)$; here, we used the notation: 
\[
\hat{\theta _{1}}(r)=\tan ^{-1}\left( \frac{2n_{1}\sqrt{\tilde{F}_{E}(r)}}{%
\sqrt{r^{2}+n_{1}^{2}}}\right) .
\]%
As in the four-dimensional case \cite{nuttybubbles} these limits will
describe an `ergocone' for the space-time. The above angle vanishes at $%
r=r_{0}$ and at infinity, while it attains a maximum value in between.

The bolt solution corresponds to a four-dimensional fixed-point set of the
isometry generated by $\partial/\partial\chi$. By solving the above
constraint on the possible periodicities of the $\chi$ coordinate we obtain
the location of the bolt $r_0=2n_2/k$. Requiring that $r\geq r_0>n_2$
implies $k=1$, in which case the periodicity of the $\chi$ coordinate is $%
8\pi n_2$ and the mass parameter is $m=n_2(15n_1^2+4n_2^2)/12$. Notice that
in this case the mass parameter is positive. The induced geometry on the
`bubblebolt' is a two-dimensional de Sitter space times a sphere $S^2$: 
\begin{eqnarray}
ds_2^2&=&(n_1^2+4n_2^2)(-dt^2+\cosh^2td\phi_1^2)+3n_2^2(d\theta_2^2+\sin^2
\theta_2 d\phi_2^2),  \nonumber \\
ds_2^2&=&(n_1^2+4n_2^2)(-dt^2+\sinh^2td\phi_1^2)+3n_2^2(d\theta_2^2+\sin^2
\theta_2 d\phi_2^2),  \nonumber \\
ds_2^2&=&(n_1^2+4n_2^2)(-dt^2+e^{2t}d\phi_1^2)+3n_2^2(d\theta_2^2+\sin^2
\theta_2 d\phi_2^2),  \nonumber \\
ds_2^2&=&(n_1^2+4n_2^2)(d\theta_1^2-\sin^2\theta_1
dt^2)+3n_2^2(d\theta_2^2+\sin^2\theta_2 d\phi_2^2).
\end{eqnarray}
At any fixed time, $r=2n_2$ is the `bubble-bolt', which is topologically $%
S^1\times S^2$. The $S^2$ factor is described by the $(\theta_2,\phi_2)$
coordinates and it has constant size in time. On the other hand, the circle $%
S^1$ described by the $\phi_1$ coordinate expands or contracts in time.
Again, the first three geometries describe three different evolutions of
this circle. The last geometry is static and it is easy to see that it
possesses an ergocone with qualitatively the same features as described
above for the static bubblenut ergocone.

We can also consider Taub-NUT spaces for which both the $2$-dimensional
factors $M_i$ are taken to be both a torus $T^2$ or a hyperboloid $H^2$.
Such geometries and the nutty bubbles obtained from them are presented in
Appendix A.

Finally, let us notice that all the nutty bubbles geometries exhibited so
far have no curvature singularities. Generically, from the form of the
metrics one would expect that $r=n_{2}$ be a curvature singularity. However,
the bubblenut solution described above is completely regular at $r=n_{2}$ as
one can check by looking at some of the curvature invariants (for example  $%
R_{\alpha \beta \gamma \delta }R^{\alpha \beta \gamma \delta }$). For the
bubble-bolt, this curvature singularity is simply avoided by requiring that $%
r\geq 2n_{2}$.

\subsection{Bubbles in cosmological backgrounds}

Another class of solutions is given for base spaces that are products of $2$%
-dimensional Einstein manifolds $M_1\times M_2$. In this case, the metric
ansatz that we use to construct the Taub-NUT solution is the one given in (%
\ref{l2}), where now $M=M_1$ while $Y=M_2$.

As an example we shall consider again the case in which $M_{1}=M_{2}=S^{2}$.
The metric is written in the form \cite{csrm}: 
\begin{eqnarray}
ds^{2} &=&-F(r)(dt-2n\cos \theta _{1}d\phi _{1})^{2}+F^{-1}(r)dr^{2} 
\nonumber \\
&&+(r^{2}+n^{2})(d\theta _{1}^{2}+\sin ^{2}\theta _{1}d\phi _{1}^{2})+\alpha
r^{2}(d\theta _{2}^{2}+\sin ^{2}\theta _{2}d\phi _{2}^{2}).  \label{619}
\end{eqnarray}%
In order to satisfy the field equations we must take 
\[
\alpha =\frac{2}{2-\lambda n^{2}},~~~~~F(r)=\frac{%
3r^{5}+(l^{2}+10n^{2})r^{3}+3n^{2}(l^{2}+5n^{2})r-6ml^{2}}{%
3rl^{2}(r^{2}+n^{2})}.\label{6dbubblecosmF} 
\]%
The metric (\ref{619}) is a solution of the vacuum Einstein field equations
with cosmological constant $\lambda =-\frac{10}{l^{2}}$, for any values of $%
n $ or $\lambda $. However, in order retain a metric of Lorentzian signature
we must ensure that $\alpha >0$, which translates in our case to $\lambda
n^{2}<2$. For convenience, we have given above the form of $F(r)$ using a
negative cosmological constant and in this case the constraint on $n$ and $%
\lambda $ is superfluous. We can also use a positive cosmological constant
(we have to analytically continue $l\rightarrow il$ in $F(r)$) and as long
as the above condition on $\alpha$ is satisfied the final metric has
Lorentzian signature. The Euclidian section is 
\begin{eqnarray}
ds^{2} &=&F_{E}(r)(d\chi +2n\cos \theta _{1}d\phi
_{1})^{2}+F_{E}^{-1}(r)dr^{2}  \nonumber \\
&&+(r^{2}-n^{2})(d\theta _{1}^{2}+\sin ^{2}\theta _{1}d\phi _{1}^{2})+\alpha
_{E}r^{2}(d\theta _{2}^{2}+\sin ^{2}\theta _{2}d\phi _{2}^{2})  \nonumber \\
\alpha _{E} &=&\frac{l^{2}}{l^{2}-5n^{2}},~~~~~~F_{E}(r)=\frac{%
3r^{5}+(l^{2}-10n^{2})r^{3}-3n^{2}(l^{2}-5n^{2})r-6ml^{2}}{%
3rl^{2}(r^{2}-n^{2})}
\end{eqnarray}%
obtained by continuing $t\rightarrow i\chi $ and $n\rightarrow in$.

We are now ready to construct the nutty bubbles. First, let us notice that
we can analytically continue the coordinates independently in the two $S^{2}$
sectors. Let us perform the analytic continuation of one of the
coordinates in the second $S^{2}$ factor, in which case we obtain the
metrics: 
\begin{eqnarray}
ds^{2} &=&F_{E}(r)(d\chi +2n\cos \theta _{1}d\phi
_{1})^{2}+F_{E}^{-1}(r)dr^{2},  \nonumber \\
&&+(r^{2}-n^{2})(d\theta _{1}^{2}+\sin ^{2}\theta _{1}d\phi _{1}^{2})+\alpha
_{E}r^{2}(-dt^{2}+\cosh ^{2}td\phi _{2}^{2}),  \nonumber \\
ds^{2} &=&F_{E}(r)(d\chi +2n\cos \theta _{1}d\phi
_{1})^{2}+F_{E}^{-1}(r)dr^{2},  \nonumber \\
&&+(r^{2}-n^{2})(d\theta _{1}^{2}+\sin ^{2}\theta _{1}d\phi _{1}^{2})+\alpha
_{E}r^{2}(-dt^{2}+\sinh ^{2}td\phi _{2}^{2}),  \nonumber \\
ds^{2} &=&F_{E}(r)(d\chi +2n\cos \theta _{1}d\phi
_{1})^{2}+F_{E}^{-1}(r)dr^{2},  \nonumber \\
&&+(r^{2}-n^{2})(d\theta _{1}^{2}+\sin ^{2}\theta _{1}d\phi _{1}^{2})+\alpha
_{E}r^{2}(-dt^{2}+e^{2t}d\phi _{2}^{2}),  \nonumber \\
ds^{2} &=&F_{E}(r)(d\chi +2n\cos \theta _{1}d\phi
_{1})^{2}+F_{E}^{-1}(r)dr^{2},  \nonumber \\
&&+(r^{2}-n^{2})(d\theta _{1}^{2}+\sin ^{2}\theta _{1}d\phi _{1}^{2})+\alpha
_{E}r^{2}(d\theta _{2}^{2}-\sin ^{2}\theta _{2}dt^{2}).
\end{eqnarray}%
The above metrics are solutions of Einstein field equations with
cosmological constant for any values of $\lambda =-10/l^{2}$ and $n$. In the
case considered here, for a negative cosmological constant, $\alpha _{E}$
can have negative values if $5n^2>l^2$. However, while in the Euclidian
sector negative values of $\alpha_{E}$ are not permitted, for our nutty
bubbles a negative value for $\alpha _{E}$ amounts to an overall sign change
of the metric in the $(t,\phi _{2})$ (respectively $(t,\theta _{2})$)
sectors. We shall see that this can have a dramatic influence on the
dynamical evolution of the bubble.

The bubble will be located at the highest root $r_0$ of $F_E(r)$ chosen such
that $r_0>n$ and in general we restrict the range of the $r$ coordinate $%
r\geq r_0$. Elimination of the Misner string sets the periodicity of the $%
\chi$ coordinate to be $8\pi n/k$ and we also have to match it with the
periodicity $4\pi/|F_E^{\prime}(r_0)|$ introduced after we eliminate any
possible conical singularities in the $(\chi,r)$ sector. Again we have two
solutions: a nut and a bolt.

The nut solution corresponds to a two-dimensional fixed-point set of the
vector $\frac{\partial }{\partial \chi }$ located at $r=n$. The periodicity
of the $\chi $ coordinate is in this case equal to $8\pi n$ and the value of
the mass parameter is fixed to $m_{n}=\frac{n^{3}(4n^{2}-l^2)}{3l^{2}}$.
Notice that the mass parameter can take any values: positive, negative or
zero. There is a curvature singularity at the bubble location! Furthermore,
if $n=\frac{l}{2}$ then curvature singularity present at $r=n$ disappears
and the spacetime has constant curvature. The fixed-points set of the
isometry generated by $\partial /\partial \chi $ is effectively two
dimensional. The induced geometry on the `bubblenut' is a two-dimensional de
Sitter space: 
\begin{eqnarray}
ds_{2}^{2} &=&\alpha _{E}n^{2}(-dt^{2}+\cosh ^{2}td\phi _{2}^{2}),  \nonumber
\\
ds_{2}^{2} &=&\alpha _{E}n^{2}(-dt^{2}+\sinh ^{2}td\phi _{2}^{2}),  \nonumber
\\
ds_{2}^{2} &=&\alpha _{E}n^{2}(-dt^{2}+e^{2t}d\phi _{2}^{2}),  \nonumber \\
ds_{2}^{2} &=&\alpha _{E}n^{2}(d\theta_2 ^{2}-\sin ^{2}\theta_2 dt^{2}).
\label{partialnutds}
\end{eqnarray}%
If the coordinate $\phi _{2}$ is periodically identified then at any fixed
time $r=n$ is our `bubblenut': a circle with minimal circumference which
expands or contracts. The first three de Sitter geometries above correspond
to three different evolutions of this circle as with (\ref{5dsbubbles}). The
last geometry is static as in these coordinates the metric has a Killing
vector $\xi =\partial /\partial t$.

Now let us consider the effect of changing the sign of $\alpha_{E}$. This
can be easily accommodated by taking $l^{2}<5n^{2}$. As we can easily see
from the metric induced on the bubble, a negative sign of $\alpha _{E}$
amounts to changing the induced de Sitter geometry of the bubblenut into a
two-dimensional anti-de Sitter geometry. As it is well known, this space can
have non-trivial identifications and so it can describe for instance a
two-dimensional black hole (as in the second and third metrics above), while
the first metric describes pure AdS in standard coordinates. After a
coordinate transformation these metrics can be written in the form: 
\begin{eqnarray}
ds^{2} &=&(-\alpha _{E})n^{2}\left( \frac{dx^{2}}{x^{2}+k}%
-(x^{2}+k)dt^{2}\right),  \nonumber \\
ds_{2}^{2} &=&(-\alpha _{E})n^{2}(-dt^{2}+\sin ^{2}td\phi _{2}^{2}),
\label{partialnutads}
\end{eqnarray}
where $k=-1,1,0$ for the respective first three metrics in (\ref%
{partialnutds}). They are all locally equivalent under changes of
coordinates. However, depending on the identifications made, the global
structure can be quite different. For example, the second metric from (\ref%
{partialnutads}) can be locally transformed into the $k=-1$ metric by a
coordinate transformation. However, if $\phi _{2}$ is periodic then the
global structure of these spaces is completely different. For $r\geq n$ this
metric describes a bubble located at $r=n$, which expands from zero size to
a finite size $(-\alpha _{E}n^{2})$ and then contracts to zero size again.
Near the initial expansion (or the final contraction) the scale factor is
linear in time and the spacetime expands or contracts like a Milne universe.

The bubble bolt geometry has a four-dimensional fixed-point set of $\frac{%
\partial}{\partial\chi}$ located at $r=r_b$ with: 
\begin{eqnarray}
r_b&=&\frac{kl^2\pm\sqrt{k^2l^4-80n^2l^2+400n^4}}{20n},
\end{eqnarray}
while the value of the mass parameter is: 
\begin{eqnarray}
m_b&=&\frac{3r_b^5+(l^2-10n^2)r_b^3-3n^2(l^2-5n^2)r_b}{6l^2}.
\end{eqnarray}
The periodicity of the coordinate $\chi$ is $\frac{8\pi n}{k}$, where $k$ is
an integer. The induced geometry on the `bubblebolt' is a two-dimensional de
Sitter space times a sphere $S^2$: 
\begin{eqnarray}
ds_2^2&=&(r_b^2-n^2)(d\theta_1^2+\sin^2\theta_1
d\phi_1^2)+\alpha_Er_b^2(-dt^2+\cosh^2td\phi_2^2),  \nonumber \\
ds_2^2&=&(r_b^2-n^2)(d\theta_1^2+\sin^2\theta_1
d\phi_1^2)+\alpha_Er_b^2(-dt^2+\sinh^2td\phi_2^2),  \nonumber \\
ds_2^2&=&(r_b^2-n^2)(d\theta_1^2+\sin^2\theta_1
d\phi_1^2)+\alpha_Er_b^2(-dt^2+e^{2t}d\phi_2^2),  \nonumber \\
ds_2^2&=&(r_b^2-n^2)(d\theta_1^2+\sin^2\theta_1
d\phi_1^2)+\alpha_Er_b^2(d\theta_2^2-\sin^2\theta_2 dt^2).
\label{6dboltcosm}
\end{eqnarray}
At any fixed time, $r=r_b$ is the `bubblebolt', which is topologically $%
S^1\times S^2$. The $S^2$ factor is described by the $(\theta_1,\phi_1)$
coordinates and it has constant size in time. On the other hand, the circle $%
S^1$ described by $\phi_2$, expands or contracts in time. Again, the first
three geometries describe three different evolutions of this circle. The
last geometry in (\ref{6dboltcosm}) is static. As for the bubblenut,
changing the sign of $\alpha_E$ has dramatic consequences as it effectively
turns the two-dimensional $dS$ geometry into $AdS$.

The boundary geometry for these bubble spacetimes is given by 
\begin{eqnarray}
ds^2&=&4n^2/l^2(d\tilde{\chi}+\cos\theta_1d\phi_1)^2+(d\theta_1^2+\sin^2%
\theta_1d\phi_1^2)+\alpha_EdS_2,  \nonumber \\
ds^2&=&4n^2/l^2(d\tilde{\chi}+\cos\theta_1d\phi_1)^2+(d\theta_1^2+\sin^2%
\theta_1d\phi_1^2)+(-\alpha_E)d\Sigma_2,
\end{eqnarray}
where $\tilde{\chi}=\chi/2n$ is periodic with period $4\pi$. Here $dS_2$
(respectively $d\Sigma_2$) describes the metric of a two-dimensional de
Sitter space (respectively $AdS$). The $(\chi,\theta_1,\phi_1)$-sector
describes a squashed three-sphere, the squashing parameter being controlled
by $4n^2/l^2$. It is interesting to note that, for negative $\alpha_E$, one
can perform identifications on the $AdS$ part of the metric which turn it
into a black-hole.

Finally, the other possibility to obtain bubble spacetimes is to
analytically continue $t\rightarrow i\chi $ and one of the coordinates in
the first $S^{2}$ factor in (\ref{619}): 
\begin{eqnarray}
ds^{2} &=&F(r)(d\chi +2n\sinh td\phi _{1})^{2}+F^{-1}(r)dr^{2}  \nonumber \\
&&+(r^{2}+n^{2})(-dt^{2}+\cosh ^{2}td\phi _{1}^{2})+\alpha r^{2}(d\theta
_{2}^{2}+\sin ^{2}\theta _{2}d\phi _{2}^{2}),  \nonumber \\
ds^{2} &=&F(r)(d\chi +2n\cosh td\phi _{1})^{2}+F^{-1}(r)dr^{2}  \nonumber \\
&&+(r^{2}+n^{2})(-dt^{2}+\sinh ^{2}td\phi _{1}^{2})+\alpha r^{2}(d\theta
_{2}^{2}+\sin ^{2}\theta _{2}d\phi _{2}^{2}),  \nonumber \\
ds^{2} &=&F(r)(d\chi +2ne^{t}d\phi _{1})^{2}+F^{-1}(r)dr^{2},  \nonumber \\
&&+(r^{2}+n^{2})(-dt^{2}+e^{2t}d\phi _{1}^{2})+\alpha r^{2}(d\theta
_{2}^{2}+\sin ^{2}\theta _{2}d\phi _{2}^{2}),  \nonumber \\
ds^{2} &=&F(r)(d\chi +2n\cos \theta _{1}dt)^{2}+F^{-1}(r)dr^{2},  \nonumber
\\
&&+(r^{2}+n^{2})(d\theta _{1}^{2}-\sin ^{2}\theta _{1}dt^{2})+\alpha
r^{2}(d\theta _{2}^{2}+\sin ^{2}\theta _{2}d\phi _{2}^{2}).
\end{eqnarray}%
The above metrics are solutions of vacuum Einstein field equations with
cosmological constant for any values of $n$ or $\lambda =-10/l^{2}$.
However, in order to keep the signature of the metric Lorentzian we have to
ensure that $\alpha >0$ i.e. $\lambda n^{2}<2$. We can have a positive%
\footnote{%
To write the solution for a positive cosmological constant we have to send $%
l\rightarrow il$ in $F(r)$ in (\ref{6dbubblecosmF}).} or negative
cosmological constant as long as this relation is satisfied. Notice that for
a negative cosmological constant $\alpha $ is always positive. The bubble is
located at the biggest root $r_{0}$ of $F(r)$ and in order to eliminate a
conical singularity in the $(\chi ,r)$-sector, we have to periodically
identify $\chi $ with period given by $4\pi /|F^{\prime }(r_{0})|=4\pi
l^{2}r_{0}/[l^{2}+5(r_{0}^{2}+n^{2})]$. At any fixed time, $r=r_{0}$ is the
bubble, which is topologically $S^{1}\times S^{2}$. The $S^{2}$ factor is
described by the $(\theta _{2},\phi _{2})$ coordinates and it has constant
size. On the other hand, the circle $S^{1}$ described by $\phi _{1}$ expands
or contracts in time. Again, the first three geometries describe three
different evolutions of this circle. The last geometry is stationary and it
is easy to see that it possesses an ergocone with qualitatively the same
features as described above for the static bubblenut ergocones encountered
in the previous sections.

Similar nutty bubbles can be obtained by considering Taub-NUT metrics for
which $M_{1}\neq M_{2}$. Such metrics have been studied in \cite%
{csrm,Page1,TNletter}. In six dimensions, such metrics can have two
independent nut parameters and there exist a constraint on the values of
these nut parameters and the cosmological constant. Quite generically this
constraint makes it impossible to set the cosmological constant to zero.
Having two nut parameters at our disposal it is very easy to construct
various nutty bubble solutions by analytically continuing the coordinates in
only one of the factors $M_{i}$. We exhibit in Appendix A more examples of
such geometries.

\subsection{Nutty bubbles and Hopf dualities}

We shall now briefly describe a method to generate new time-dependent
solutions starting from some of the nutty bubbles studied in the previous
sections. This method was based on the fact that, in general, the
odd-dimensional spheres $S^{2n+1}$ may be regarded as circle bundles over $%
CP^{n}$ and one can use the so-called Hopf duality (a T-duality along the $%
U(1)$-fibre) to generate new solutions by untwisting $S^{2n+1}$ to $%
CP^{n}\times S^{1}$, as in \cite{Duff1,Duff2,Cvetic,KKM}. The
six-dimensional case is particularly interesting for us since it has been
shown in \cite{Duff2} that it is possible to make consistent truncations of
the maximal Type II supergravity theories to a bosonic sector which exhibits
an $O(2,2)$ global symmetry, with the $T$-duality transformation taking a
very simple form. The theories at hand are the toroidal reductions of Type
IIA, respectively Type IIB ten-dimensional supergravities, while the
reduction ansatz for the fields is that the six-dimensional fields that are
retained are precisely the ten-dimensional ones, with the spacetime indices
restricted to run over the six-dimensional range only. The two truncated
theories in $D=6$ are then related by a T-duality transformation upon
reduction to $D=5$. The explicit mappings of the fields have been given in %
\cite{Duff2} and we mainly follow their notational conventions. However, for convenience we also provide the derivation of the $T$-duality rules
in Appendix B.

In this section we apply Hopf dualities to some of our nutty bubbles to
generate new time dependent backgrounds. However, since these rules work only for
solutions that do not have cosmological constant, we shall focus
mainly on the cases in which $M_1=M_2$. For simplicity, and just to
illustrate the method we shall use just a few nutty bubbles solutions as
seeds.

Let us start with the solution given in (\ref{tildes2}). Considering this
metric as a solution of the pure gravity sector of the truncated Type IIA
theory we can now perform a Hopf-duality along the spacelike $\chi$%
-direction to obtain a solution of six-dimensional Type IIB theory: 
\begin{eqnarray}
ds_{6B} &=&\tilde{F}_E(r)^{-\frac{1}{2}}d\chi^{2}+\tilde{F}_E(r)^{-\frac{1}{2%
}}dr^{2}+\tilde{F}_E(r)^{\frac{1}{2}}(r^{2}+n_{1}^{2})(-dt^{2}+\cosh^{2}td
\phi_1^{2})  \nonumber \\
&&+\tilde{F}_E(r)^{\frac{1}{2}}(r^{2}-n_{2}^{2})(d\theta _{2}^{2}+\sin
^{2}\theta _{2}d\phi_2^{2})^{2}  \nonumber \\
e^{2\varphi _{1}} &=&e^{2\varphi _{2}}=\tilde{F}_E(r),~~~~~~~
A_{(2)}^{NS}=2n_{1}\sinh t d\phi _{1}\wedge d\chi+2n_{2}\cos \theta
_{2}d\phi_2\wedge d\chi.
\end{eqnarray}

Were we to consider (\ref{tildes2}) as a solution of the pure gravity sector
of Type IIB theory, then after performing the spacelike Hopf dualisation we
would obtain a solution of Type IIA theory: 
\begin{eqnarray}
ds_{6A} &=&\tilde{F}(r)^{-\frac{1}{2}}d\chi ^{2}+\tilde{F}(r)^{-\frac{1}{2}%
}dr^{2}+\bar{F}(r)^{\frac{1}{2}}(r^{2}+n_{1}^{2})(-dt^{2}+\cosh ^{2}td\phi_1
^{2})  \nonumber \\
&&+\bar{F}(r)^{\frac{1}{2}}(r^{2}-n_{2}^{2})(d\theta _{2}^{2}+\sin
^{2}\theta _{2}d\phi_2 ^{2})  \nonumber \\
e^{2\varphi _{1}} &=&e^{2\varphi _{2}}=\bar{F}(r),~~~~~~~A_{(2)}=2n_{1}\sinh
td\phi _{1}\wedge d\chi +2n_{2}\cos \theta _{2}d\phi _{2}\wedge d\chi.
\end{eqnarray}

The analysis of these charged bubbles proceeds as in the previous sections.
The bubble will be located at the largest root of $\tilde{F}_{E}(r)$.
Generically there exists a curvature singularity at the bubble location,
which cannot be cured by any appropriate choices of the parameters. Another
difference with the previous bubble solutions is that in the $(\chi ,r)$%
-sector there is no conical singularity to be eliminated and $\chi $ need
not be compactified.

As another example of this method, let us consider a bubble solution, given
in Appendix A, which corresponds to a six-dimensional Taub-NUT constructed
as a circle fibration over $T^{2}\times T^{2}$. Taking the bubble solution
given in (\ref{T2xT2}) as a solution of Type IIA theory, then after
performing a Hopf duality along the $\chi $ direction we obtain: 
\begin{eqnarray}
ds_{6B} &=&\tilde{F}_{E}(r)^{-\frac{1}{2}}d\chi ^{2}+\tilde{F}_{E}(r)^{-%
\frac{1}{2}}dr^{2}+\tilde{F}_{E}(r)^{\frac{1}{2}}(r^{2}+n_{1}^{2})(-dt^{2}+d%
\phi _{1}^{2})+\tilde{F}_{E}(r)^{\frac{1}{2}}(r^{2}-n_{2}^{2})(d%
\theta_{2}^{2}+d\phi _{2}^{2})  \nonumber \\
e^{2\varphi _{1}} &=&e^{2\varphi _{2}}=\tilde{F}%
_{E}(r),~~~~~~~A_{(2)}^{NS}=2n_{1}td\phi _{1}\wedge d\chi +2n_{2}\theta
_{2}d\phi _{2}\wedge d\chi.
\end{eqnarray}
which is a solution of the Type IIB theory. Notice that $F(r)=0$ only if $%
r=0 $. On the other hand, we have to restrict the values of the radial
coordinates such that $r\geq n_{2}$, or else the signature of this metric
will change. There is however a curvature singularity at $r=n_{2}$, which
cannot be eliminated by any appropriate choices of the parameter $m$.

\section{Discussion}

In this paper, we have constructed a wide variety of time-dependent
backgrounds using the standard techniques of analytic continuation. Since
many of the presented solutions are locally asymptotically (A)dS, they are
relevant in the context of gauge/gravity dualities.

For example, let us discuss one of our solutions (\ref{5dsbubbles}) in the
context of the AdS/CFT correspondence. The bulk-boundary correspondence in
the Lorentzian section demands the inclusion of both normalizable and
non-normalizable modes of the bulk fields \cite{adscft}. The former propagate 
in the bulk and
correspond to physical states while the latter serve as classical,
non-fluctuating backgrounds and encode the choice of operator insertions in
the boundary theory.

Since the bulk theory is a theory of gravity, one of the bulk fields will
always be the graviton (metric perturbations). The AdS/CFT dictionary tells
us that its dual operator is the stress-energy tensor of the CFT. We will
compare the dual CFT stress tensor to the $rescaled$ boundary stress tensor
calculated from the bulk spacetime using the counterterm subtraction
procedure of~\cite{balkraus,kostas}. Typically, the boundary of a locally
asymptotically spacetime will be an asymptotic surface at some large radius $%
r$. However, the metric restricted to the boundary $\gamma_{ab}$ diverges
due to an infinite conformal factor $r^{2}/\ell ^{2}$, and so the metric
upon which the dual field theory resides is usually defined using the
rescaling 
\begin{eqnarray}  \label{resc}
h_{ab}=\lim_{r\rightarrow \infty }\frac{\ell ^{2}}{r^{2}}\gamma _{ab}.
\end{eqnarray}%
Corresponding to the boundary metric $h_{ab}$, the stress-energy tensor $%
<\tau _{ab}>$ for the dual theory can be calculated using the following
relation 
\begin{eqnarray}  \label{r1}
\sqrt{-h}h^{ab}<\tau _{bc}>=\lim_{r\rightarrow \infty }\sqrt{-\gamma }\gamma
^{ab}T_{bc}.
\end{eqnarray}
In our case, the boundary metric is 
\begin{eqnarray}
ds^2=h_{ab}dx^adx^b&=&(d\chi+lxdt)^2+l^2\left(\frac{dx^2}{x^2+k}%
-(x^2+k)dt^2\right)+l^2dy^2,
\end{eqnarray}
and so the conformal boundary, where the $\mathcal{N}=4$ SYM lives, is $%
AdS_3\times S^1$. The $rescaled$ boundary stress tensor is 
\begin{eqnarray}  \label{st}
\tau^t_t&=&\frac{256m+5l^2}{1024\pi Gl^3} ,  \nonumber \\
\tau^t_{\chi}&=&0,  \nonumber \\
\tau^{\chi}_{t}&=&-\frac{(l^2+64m)x}{64\pi Gl^3}  \nonumber \\
\tau^{\chi}_{\chi}&=&-\frac{11l^2+768m}{1024\pi Gl^3},  \nonumber \\
\tau^x_x&=&\frac{5l^2+256m}{1024\pi Gl^3},  \nonumber \\
\tau^y_y&=&\frac{l^2+256m}{1024\pi Gl^3}.
\end{eqnarray}
Since the boundary metric is the product of a circle and a three-dimensional
Einstein space, the trace anomaly vanishes. Indeed, as we expected, the
stress tensor (\ref{st}) is finite, covariantly conserved, and manifestly
traceless.

For four dimensions, it was shown in \cite{nuttybubbles} that in the special
case when the NUT charge vanishes ($n=0$), the metric (stress tensor)
reduces to the 4-dimensional Schwarzschild-AdS metric (stress tensor). In
five dimensions, the constraint between the NUT charge and the cosmological
constant changes dramatically the situation --- we will comment on this at
the end of this section. However, the limit we are interested here is $%
m=-64/l^2$. Then, the bulk geometry has constant curvature and it is the
static bubble obtained from $AdS_5$ by analytic continuation. Indeed, in
this case $F(r)=\frac{r^2}{l^2}+\frac{1}{4}$ and by redefining the
coordinate $r^2\rightarrow r^2-\frac{l^2}{4}$ and rescaling $y$ to absorb an 
$l^2$ factor, the metric can be cast in the form: 
\begin{eqnarray}
ds^2&=&\left(\frac{r^2}{l^2}-\frac{1}{4}\right)dy^2+\frac{dr^2}{\left(\frac{%
r^2}{l^2}-\frac{1}{4}\right)}+r^2\big[(d\tilde{\chi}+\sinh\theta
dt)^2+d\theta^2-\cosh^2\theta dt^2\big].  \nonumber
\end{eqnarray}
One can recognize it as being the analytic continuation of $AdS_5$ with a
non-canonically normalized $H^3$ factor. For this particular value of the
parameter $m$, the stress tensor (\ref{st}) becomes 
\begin{eqnarray}  \label{st1}
\tau^a_b = \frac{N^2}{512 \pi^2 l^4} \mbox{
diag}(1,1,1,-3),
\end{eqnarray}
where we have used the standard relation $l^3/G = 2N^2/\pi$ to rewrite the
stress tensor in terms of field theory quantities.\footnote{%
The convention for the coordinates is ${1,2,3,4}={t, \chi, x, y}$.}

Let us move now to the dual theory that is in terms of $\mathcal{N}=4$ SYM
on the $AdS_3 \times S^1$ spacetime. This is a conformally flat spacetime
and, fortunately, there is a standard result for the stress tensor~\cite%
{birdav}: 
\begin{equation}  \label{ftstress}
\langle \tau^a_b \rangle = -{\frac{1 }{16 \pi^2}} \left( A {}^{(1)} H^a_b +
B {}^{(3)} H^a_b \right) + \tilde{\tau}^a_b.
\end{equation}
Here, $^{(1)} H^a_b$ and $^{(3)} H^a_b$ are conserved quantities constructed
from the curvature (see~\cite{birdav} for their definitions), and $\tilde{%
\tau}^a_b$ is a traceless state-dependent part. In our case they are given
by 
\begin{eqnarray}
^{(1)} H^a_b&=&\frac{3}{8l^4}\mbox{diag} [1,1,1,-3],  \nonumber \\
^{(3)} H^a_b&=&-\frac{1}{16l^4}\mbox{diag} [1,1,1,-3].
\end{eqnarray}
The coefficients $A$ and $B$ are calculated as in \cite%
{Balasubramanian:2002am}. The trace of (\ref{ftstress}) is compared with the
conformal anomaly for $\mathcal{N}=4$ SYM~\cite{kostas}: 
\begin{equation}
\langle \tau^a_a\rangle = -{\frac{1 }{16 \pi^2}} \left( -6A \Box R - B
(R_{ab} R^{ab} - 1/3 R^2) \right) = {\frac{(N^2-1) }{64 \pi^2}} (2 R_{ab}
R^{ab} - 2/3 R^2).
\end{equation}
This fixes $A=0$ and $B=(N^2-1)/2$ and so the field theory stress tensor
becomes 
\begin{equation}  \label{st2}
\langle \tau^a_b \rangle = \frac{1}{2}{\frac{(N^2-1) }{256 \pi^2 l^4}} %
\mbox{ diag}(1,1,1,-3) + \tilde{\tau}^a_b.
\end{equation}
In the large $N$ limit, the geometrical part of the stress tensor precisely
reproduces (\ref{st1}). The fact that the geometrical part is non zero is a
direct consequence of analytic continuation --- the quantum field theory on
the AdS boundary can have a nonvanishing vacuum (Casimir) energy.
Consequently, the above comparison of the stress tensor (\ref{st1}) to (\ref%
{st2}) does result in a non-trivial connection between them.

As advertised in Section $2$, we would like to comment here on the limit
when the NUT charge and/or the cosmological constant are zero. Central to
our construction was starting with an appropriate NUT-charged family of
(generating) solutions in higher dimensions. These solutions can have more
than one nut parameter. Since in higher dimensions there is a constraint
between the nut parameters and the cosmological constant, the limit
mentioned above is more subtle than in four dimensions. Of course, these
remarks apply to odd dimensions, particularly to the $5$-dimensional
solutions in Section $2$. However, as noted there, there is a way to evade
this situation: after a change of coordinates and by performing appropriate
rescalings, it turns out that that the metric can be cast in such a form that
allows us to take the limit of a vanishing cosmological constant.\footnote{%
The metric obtained this way is a generalization of the Eguchi-Hanson metric
in higher dimensions \cite{rick} (see, also, \cite{TNletter}).}

More precisely, start with the metric
\begin{equation}
ds^{2}=F_{E}(r)(d\chi -l\cos \theta d\varphi
)^{2}+F_{E}^{-1}(r)dr^{2}+\left( r^{2}-\frac{l^{2}}{4}\right) (d\theta
^{2}+\sin ^{2}\theta d\varphi ^{2})+r^{2}dz^{2},
\end{equation}%
where 
\begin{equation}
F_{E}(r)=\frac{4r^{4}-2l^{2}r^{2}+16ml^{2}}{l^{2}(4r^{2}-l^{2})}.
\end{equation}%
Make now the coordinate change $r^{2}\rightarrow r^{2}+l^{2}/4$ and define $%
a^{4}=64ml^{2}-l^{4}$. Then, after a further rescaling of the $r$ and $y$
coordinates the metric becomes 
\begin{eqnarray}
ds^{2} &=&\frac{r^{2}}{4}\left( 1-\frac{a^{4}}{r^{4}}\right) (d\tilde{\chi}%
+\cos \theta d\phi )^{2}+\frac{dr^{2}}{\left( \frac{r^{2}}{l^{2}}-1\right)
\left( 1-\frac{a^{4}}{r^{4}}\right) }  \nonumber \\
&&+\frac{r^{2}}{4}(d\theta ^{2}+\sin ^{2}\theta d\phi ^{2})+\left( \frac{%
r^{2}}{l^{2}}-1\right) d\tilde{y}^{2},
\end{eqnarray}%
which is a solution of the $5$-dimensional Einstein
equations with cosmological constant, referred to as the Eguchi-Hanson
soliton \cite{rick}. The limit in which the cosmological constant vanishes
is now a smooth limit and the metric becomes the product of the
four-dimensional Eguchi-Hanson metric with a trivial flat direction.

There also exists a limit in which we can set the nut parameter to zero.
Namely, our $5$-dimensional metric can be written quite generally in the
form \cite{TNletter} 
\begin{eqnarray}
ds^2&=&-F(r)(dt+2\frac{n}{\delta}\cos\theta d\phi)^2+\frac{dr^2}{F(r)}+\frac{%
r^2+n^2}{\delta} (d\theta^2+\sin^2\theta d\phi^2)+r^2dy^2, \nonumber \\
F(r)&=&\frac{r^4+2n^2r^2-2ml^2}{r^2+n^2},
\end{eqnarray}
and the constrain equation is now simply $\delta=-\frac{4n^2}{l^2}$. This
metric is then a solution of Einstein field equations with cosmological
constant $\Lambda=-\frac{6}{l^2}$. Once we fix $\delta$ as above, there is no
constraint on the values of $\Lambda$ and $n$ other than the requirement of
a metric of Lorentzian signature --- this can be easily accommodated by
analytically continuing the coordinate $\theta\rightarrow i\theta$. Defining
now a new nut parameter $N=\frac{n}{\delta}$ and $\lambda=-\frac{4}{l^2}$, 
the above solution can be written in the following form: 
\begin{eqnarray}
ds^2&=&-F(r)(dt-2N\cosh\theta d\phi)^2+\frac{dr^2}{F(r)}+\frac{\lambda^2
N^2r^2+1} {(-\lambda)}(d\theta^2+\sinh^2\theta d\phi^2)+r^2dy^2,  \nonumber \\
F(r)&=&\frac{16N^2r^4+2r^2l^2-ml^2}{l^2(16N^2r^2+l^4)}.
\end{eqnarray}
When $N\neq 0$, a change of coordinates will bring the metric into a form
similar to the one discussed in Section $2$. Notice however that, in the
form written above, it is possible to take a smooth limit of the metric in
which the nut charge $N\rightarrow 0$. Then, we obtain a metric
that is the trivial product of a $3$-dimensional Schwarzschild AdS (described by the coordinates $(t,r,y)$) with a $2$-dimensional hyperboloid
(described by $(\theta, \phi)$). 

Finally, we note that the boundary of the solutions presented in the paper 
is generically a circle-fibration over base spaces --- it is obtained from products of
general Einstein-K\"ahler manifolds and can have exotic topologies.
Then, one should be able to understand the thermodynamic phase structure of
the dual field theory by working out the corresponding phase structure of
our gravity solutions in the bulk. We leave a more detailed study of these
solutions and their duals for future work.

\medskip

{\Large Acknowledgements}

DA was supported by the Department of Atomic Physics, Government of India
and the visitor programme of Perimeter Institute. RBM and CS were supported
by the Natural Sciences \& Engineering Research Council of Canada.

\renewcommand{\theequation}{A-\arabic{equation}} 
\setcounter{equation}{0} 

\section*{Appendix A: Other six-dimensional bubble spacetimes}

\label{otherbubbles}

\subsubsection*{$U^1$-fibration over $S^2\times T^2$}

Let us consider now the case of a Taub-NUT space that appears as a radial
extension of a $U(1)$-fibration over $S^{2}\times T^{2}$ in the Euclidian
sector \cite{csrm}: 
\begin{eqnarray}
ds^{2} &=&F_{E}(r)(d\chi +2n_{1}\cos \theta _{1}d\phi _{1}+2n_{2}\theta
_{2}d\phi _{2})^{2}+F_{E}^{-1}(r)dr^{2}  \nonumber \\
&&+(r^{2}-n_{1}^{2})(d\theta _{1}^{2}+\sin ^{2}\theta _{1}d\phi
_{1}^{2})+(r^{2}-n_{2}^{2})(d\theta _{2}^{2}+d\phi _{2}^{2})  \nonumber \\
F_{E}(r) &=&\frac{%
3r^{6}+(l^{2}-5n_{2}^{2}-10n_{1}^{2})r^{4}+3(-n_{2}^{2}l^{2}+10n_{1}^{2}n_{2}^{2}-n_{1}^{2}l^{2}+5n_{1}^{4})r^{2}%
}{3(r^{2}-n_{1}^{2})(r^{2}-n_{2}^{2})l^{2}}  \nonumber \\
&&+\frac{6ml^{2}r-3n_{1}^{2}n_{2}^{2}(l^{2}-5n_{1}^{2})}{%
3(r^{2}-n_{1}^{2})(r^{2}-n_{2}^{2})l^{2}}
\end{eqnarray}%
The above metric will be solution of the Einstein field equations with
cosmological constant if and only if we have $(n_{2}^{2}-n_{1}^{2})\lambda =2
$. As a consequence we are forced to consider a non-zero cosmological
constant. We have now two possibilities: we can perform analytic
continuations in the $S^{2}$ sector or we can do that in the $T^{2}$ sector.
In the first case we make analytic continuations in the $S^{2}$ sector,
which in turn forces us to take $n_{1}\rightarrow in_{1}$: 
\begin{eqnarray}
ds^{2} &=&\tilde{F}_{E}(r)(d\chi +2n_{1}\sinh td\phi _{1}+2n_{2}\theta
_{2}d\phi _{2})^{2}+\tilde{F}_{E}^{-1}(r)dr^{2}  \nonumber \\
&&+(r^{2}+n_{1}^{2})(-dt^{2}+\cosh ^{2}td\phi
_{1}^{2})+(r^{2}-n_{2}^{2})(d\theta _{2}^{2}+d\phi _{2}^{2})  \nonumber \\
ds^{2} &=&\tilde{F}_{E}(r)(d\chi +2n_{1}\cosh td\phi _{1}+2n_{2}\theta
_{2}d\phi _{2})^{2}+\tilde{F}_{E}^{-1}(r)dr^{2}  \nonumber \\
&&+(r^{2}+n_{1}^{2})(-dt^{2}+\sinh ^{2}td\phi
_{1}^{2})+(r^{2}-n_{2}^{2})(d\theta _{2}^{2}+d\phi _{2}^{2})  \nonumber \\
ds^{2} &=&\tilde{F}_{E}(r)(d\chi +2n_{1}e^{t}d\phi _{1}+2n_{2}\theta
_{2}d\phi _{2})^{2}+\tilde{F}_{E}^{-1}(r)dr^{2}  \nonumber \\
&&+(r^{2}+n_{1}^{2})(-dt^{2}+e^{2t}d\phi _{1}^{2})+(r^{2}-n_{2}^{2})(d\theta
_{2}^{2}+d\phi _{2}^{2})  \nonumber \\
ds^{2} &=&\tilde{F}_{E}(r)(d\chi +2n_{1}\cos \theta _{1}dt+2n_{2}\theta
_{2}d\phi _{2})^{2}+\tilde{F}_{E}^{-1}(r)dr^{2}  \nonumber \\
&&+(r^{2}+n_{1}^{2})(d\theta _{1}^{2}-\sin ^{2}\theta
_{1}dt^{2})+(r^{2}-n_{2}^{2})(d\theta _{2}^{2}+d\phi _{2}^{2})  \nonumber \\
\tilde{F}_{E}(r) &=&-\frac{%
3r^{6}+(-l^{2}-5n_{2}^{2}+10n_{1}^{2})r^{4}+3(n_{2}^{2}l^{2}-10n_{1}^{2}n_{2}^{2}-n_{1}^{2}l^{2}+5n_{1}^{4})r^{2}%
}{3(r^{2}+n_{1}^{2})(r^{2}-n_{2}^{2})l^{2}}  \nonumber \\
&&+\frac{6ml^{2}r+3n_{1}^{2}n_{2}^{2}(l^{2}-5n_{1}^{2})}{%
3(r^{2}+n_{1}^{2})(r^{2}-n_{2}^{2})l^{2}}
\end{eqnarray}%
The above metrics will be solutions of the Einstein field equations with
cosmological constant if and only if $(n_{1}^{2}+n_{2}^{2})\lambda =2$.
Hence we are also constrained to have only a positive cosmological constant
which implies that our solutions are time-dependent asymptotically de Sitter
spaces (notice that, in view of this fact we have already continued $%
l\rightarrow il$ in the above expression for $\tilde{F}_{E}$). On the other
hand, if we perform analytic continuations in the $T^{2}$ sector we are
forced to take $n_{2}\rightarrow in_{2}$ and we obtain the metrics: 
\begin{eqnarray}
ds^{2} &=&\bar{F}_{E}(r)(d\chi +2n_{1}\cos \theta _{1}d\phi
_{1}+2n_{2}td\phi _{2})^{2}+\bar{F}_{E}^{-1}(r)dr^{2}  \nonumber \\
&&+(r^{2}-n_{1}^{2})(d\theta _{1}^{2}+\sin ^{2}\theta _{1}d\phi
_{1}^{2})+(r^{2}+n_{2}^{2})(-dt^{2}+d\phi _{2}^{2})  \nonumber \\
ds^{2} &=&\bar{F}_{E}(r)(d\chi +2n_{1}\cos \theta _{1}d\phi
_{1}+2n_{2}\theta _{2}dt)^{2}+\bar{F}_{E}^{-1}(r)dr^{2}  \nonumber \\
&&+(r^{2}-n_{1}^{2})(d\theta _{1}^{2}+\sin ^{2}\theta _{1}d\phi
_{1}^{2})+(r^{2}+n_{2}^{2})(d\theta _{2}^{2}-dt^{2})  \nonumber \\
\bar{F}_{E}(r) &=&\frac{%
3r^{6}+(l^{2}+5n_{2}^{2}-10n_{1}^{2})r^{4}+3(n_{2}^{2}l^{2}-10n_{1}^{2}n_{2}^{2}-n_{1}^{2}l^{2}+5n_{1}^{4})r^{2}%
}{3(r^{2}-n_{1}^{2})(r^{2}+n_{2}^{2})l^{2}}  \nonumber \\
&&+\frac{6ml^{2}r+3n_{1}^{2}n_{2}^{2}(l^{2}-5n_{1}^{2})}{%
3(r^{2}-n_{1}^{2})(r^{2}+n_{2}^{2})l^{2}}
\end{eqnarray}%
The above metrics will be solutions of Einstein field equations if and only
if $(n_{1}^{2}+n_{2}^{2})\lambda =-1$, which means that the cosmological
constant must be negative. In conclusions the above solutions are
time-dependent backgrounds that are asymptotically $AdS$.

\subsubsection*{$U(1)$-fibrations over $T^2\times T^2$}

Let us consider next the analytic continuation of the Taub-NUT spaces that
appear as $U(1)$-fibrations over $T^{2}\times T^{2}$. The Euclidean version
of such spaces is \cite{csrm}: 
\begin{eqnarray}
ds^{2} &=&F_{E}(r)(d\chi +2n_{1}\theta _{1}d\phi _{1}+2n_{2}\theta _{2}d\phi
_{2})^{2}+F_{E}^{-1}(r)dr^{2}  \nonumber \\
&&+(r^{2}-n_{1}^{2})(d\theta _{1}^{2}+d\phi
_{1}^{2})+(r^{2}-n_{2}^{2})(d\theta _{2}^{2}+d\phi _{2}^{2})  \nonumber \\
F_{E}(r) &=&\frac{%
3r^{6}-5(n_{2}^{2}+2n_{1}^{2})r^{4}+15n_{1}^{2}(n_{1}^{2}+2n_{2}^{2})r^{2}+6ml^{2}r+15n_{1}^{4}n_{2}^{2}%
}{3(r^{2}-n_{1}^{2})(r^{2}-n_{2}^{2})l^{2}}
\end{eqnarray}%
The above metric is a solution of vacuum Einstein field equations with
cosmological constant if and only if $(n_{2}^{2}-n_{1}^{2})\lambda =0$.
Hence in the case of a vanishing cosmological constant we can have two
independent NUT charges in the metric. We can similarly analytically
continue the coordinates from one factor space $T^{2}$ only: 
\begin{eqnarray}
ds^{2} &=&\tilde{F}_{E}(r)(d\chi +2n_{1}td\phi _{1}+2n_{2}\theta _{2}d\phi
_{2})^{2}+\tilde{F}_{E}^{-1}(r)dr^{2}  \nonumber \\
&&+(r^{2}+n_{1}^{2})(-dt^{2}+d\phi _{1}^{2})+(r^{2}-n_{2}^{2})(d\theta
_{2}^{2}+d\phi _{2}^{2})  \nonumber \\
ds^{2} &=&\tilde{F}_{E}(r)(d\chi +2n_{1}\theta _{1}dt+2n_{2}\theta _{2}d\phi
_{2})^{2}+\tilde{F}_{E}^{-1}(r)dr^{2}  \nonumber \\
&&+(r^{2}+n_{1}^{2})(d\theta _{1}^{2}-dt^{2})+(r^{2}-n_{2}^{2})(d\theta
_{2}^{2}+d\phi _{2}^{2})  \nonumber \\
\tilde{F}_{E}(r) &=&\frac{2mr}{(r^{2}+n_{1}^{2})(r^{2}-n_{2}^{2})}
\label{T2xT2}
\end{eqnarray}%
However, if $\lambda \neq 0$ we are forced to have $n_{1}=n_{2}$ and it is
impossible to analytically continue the coordinates of the $T^{2}$ factors
separately.

\subsection*{Warped products of nutty spaces}

These Taub-NUT spaces have the base space factorized as a product of the
form $M_{2}^{(1)}\times M_{2}^{(2)}$ where the factors $M_{2}^{(i)}$ are
two-dimensional Einstein spaces with constant curvature. We shall choose
them to be of the form $S^{2}$, $T^{2}$ or $H^{2}$. Consider now the case of
a warped product of a circle fibration over the $S^{2}$ factor of the base
space $S^{2}\times T^{2}$. The Euclidean version of the metric of these
spaces is given by: 
\begin{eqnarray}
ds^{2} &=&F_{E}(r)(d\chi +2n\cos \theta _{1}d\phi
_{1})^{2}+F_{E}^{-1}(r)dr^{2}+(r^{2}-n^{2})(d\theta _{1}^{2}+\sin ^{2}\theta
_{1}d\phi _{1}^{2})+r^{2}(d\theta _{2}^{2}+d\phi _{2}^{2})  \nonumber \\
{F_{E}}(r) &=&\frac{3r^{5}+(l^{2}+10n^{2})r^{3}-3n^{2}(l^{2}+5n^{2})r+6ml^{2}%
}{3rl^{2}(r^{2}-n^{2})}
\end{eqnarray}%
The above metric is a solution of vacuum Einstein field equations with a
cosmological constant if and only if $\lambda n^{2}=-2$, i.e. we are
constrained to have a negative cosmological constant $\lambda =-\frac{10}{%
l^{2}}$. Performing analytic continuations in the $S^{2}$ sector we are
forced to take $n\rightarrow in$ and the constraint equation will become in
this case $\lambda n^{2}=2$. Hence our solutions will be asymptotically de
Sitter: 
\begin{eqnarray}
ds^{2} &=&\tilde{F}_{E}(r)(d\chi +2n\sinh td\phi _{1})^{2}+\tilde{F}%
_{E}^{-1}(r)dr^{2}+(r^{2}+n^{2})(-dt^{2}+\cosh ^{2}td\phi
_{1}^{2})+r^{2}(d\theta _{2}^{2}+d\phi _{2}^{2})  \nonumber \\
ds^{2} &=&\tilde{F}_{E}(r)(d\chi +2n\cosh td\phi _{1})^{2}+\tilde{F}%
_{E}^{-1}(r)dr^{2}+(r^{2}+n^{2})(-dt^{2}+\sinh ^{2}td\phi
_{1}^{2})+r^{2}(d\theta _{2}^{2}+d\phi _{2}^{2})  \nonumber \\
ds^{2} &=&\tilde{F}_{E}(r)(d\chi +2ne^{t}d\phi _{1})^{2}+\tilde{F}%
_{E}^{-1}(r)dr^{2}+(r^{2}+n^{2})(-dt^{2}+e^{2t}d\phi _{1}^{2})+r^{2}(d\theta
_{2}^{2}+d\phi _{2}^{2})  \nonumber \\
ds^{2} &=&\tilde{F}_{E}(r)(d\chi +2n\cos \theta _{1}dt)^{2}+\tilde{F}%
_{E}^{-1}(r)dr^{2}+(r^{2}+n^{2})(d\theta _{1}^{2}-\sin ^{2}\theta
_{1}dt^{2})+r^{2}(d\theta _{2}^{2}+d\phi _{2}^{2})  \nonumber \\
\tilde{F}_{E} &=&\frac{%
-3r^{5}+(l^{2}-10n^{2})r^{3}+3n^{2}(l^{2}-5n^{2})r+6ml^{2}}{%
3rl^{2}(r^{2}+n^{2})}
\end{eqnarray}%
If we analytically continue the coordinates in the second sector $T^{2}$ we
obtain: 
\[
ds^{2}=F_{E}(r)(d\chi +2n\cos \theta _{1}d\phi
_{1})^{2}+F_{E}^{-1}(r)dr^{2}+(r^{2}-n^{2})(d\theta _{1}^{2}+\sin ^{2}\theta
_{1}d\phi _{1}^{2})+r^{2}(d\theta _{2}^{2}-dt^{2})
\]

We now consider partial fibrations over $S^{2}$ with the base space of the
form $S^{2}\times H^{2}$. The Euclidean version of these metrics is given
by: 
\begin{eqnarray}
ds^{2} &=&F_{E}(r)(d\chi +2n\cos \theta _{1}d\phi
_{1})^{2}+F_{E}^{-1}(r)dr^{2}+(r^{2}-n^{2})(d\theta _{1}^{2}+\sin ^{2}\theta
_{1}d\phi _{1}^{2})  \nonumber \\
&&+\alpha r^{2}(d\theta _{2}^{2}+\sinh ^{2}\theta _{2}d\phi _{2}^{2}) 
\nonumber \\
\alpha  &=&-\frac{2}{\lambda n^{2}+2},~~~~~F_{E}(r)=\frac{%
3r^{5}+(l^{2}-10n^{2})r^{3}-3n^{2}(l^{2}+5n^{2})r+6ml^{2}}{%
3rl^{2}(r^{2}-n^{2})}
\end{eqnarray}%
Since $\alpha $ has to be positive, this means that $\lambda n^{2}<-2$,
therefore the space is asymptotically anti-de Sitter. If we perform analytic
continuation in the $S^{2}$ sector we obtain: 
\begin{eqnarray}
ds^{2} &=&\tilde{F}_{E}(r)(d\chi +2n\sinh td\phi _{1})^{2}+\tilde{F}%
_{E}^{-1}(r)dr^{+}(r^{2}+n^{2})(-dt^{2}+\cosh ^{2}td\phi _{1}^{2})  \nonumber
\\
&&+\tilde{\alpha}r^{2}(d\theta _{2}^{2}+\sinh ^{2}\theta _{2}d\phi _{2}^{2})
\nonumber \\
ds^{2} &=&\tilde{F}_{E}(r)(d\chi +2n\cosh td\phi _{1})^{2}+\tilde{F}%
_{E}^{-1}(r)dr^{2}+(r^{2}+n^{2})(-dt^{2}+\sinh ^{2}td\phi _{1}^{2}) 
\nonumber \\
&&+\tilde{\alpha}r^{2}(d\theta _{2}^{2}+\sinh ^{2}\theta _{2}d\phi _{2}^{2})
\nonumber \\
ds^{2} &=&\tilde{F}_{E}(r)(d\chi +2ne^{t}d\phi _{1})^{2}+\tilde{F}%
_{E}^{-1}(r)dr^{2}+(r^{2}+n^{2})(-dt^{2}+e^{2t}d\phi _{1}^{2})  \nonumber \\
&&+\tilde{\alpha}r^{2}(d\theta _{2}^{2}+\sinh ^{2}\theta _{2}d\phi _{2}^{2})
\nonumber \\
ds^{2} &=&\tilde{F}_{E}(r)(d\chi +2n\cos \theta _{1}dt)^{2}+\tilde{F}%
_{E}^{-1}(r)dr^{2}+(r^{2}+n^{2})(d\theta _{1}^{2}-\sin ^{2}\theta _{1}dt^{2})
\nonumber \\
&&+\tilde{\alpha}r^{2}(d\theta _{2}^{2}+\sinh ^{2}\theta _{2}d\phi _{2}^{2})
\end{eqnarray}%
where 
\[
\tilde{\alpha}=\frac{2}{\lambda n^{2}-2},~~~~~~~\tilde{F}_{E}(r)=\frac{%
-3r^{5}+(l^{2}+10n^{2})r^{3}-3n^{2}(l^{2}+5n^{2})r+6ml^{2}}{%
3rl^{2}(r^{2}-n^{2})}
\]%
If we perform analytic continuations in the $H^{2}$ section of the base
space we obtain: 
\begin{eqnarray}
ds^{2} &=&F_{E}(r)(d\chi +2n\cos \theta _{1}d\phi
_{1})^{2}+F_{E}^{-1}(r)dr^{2}  \nonumber \\
&&+(r^{2}-n^{2})(d\theta _{1}^{2}+\sin ^{2}\theta _{1}d\phi _{1}^{2})+\alpha
r^{2}(d\theta _{2}^{2}-\cosh ^{2}\theta _{2}dt^{2})  \nonumber \\
ds^{2} &=&F_{E}(r)(d\chi +2n\cos \theta _{1}d\phi
_{1})^{2}+F_{E}^{-1}(r)dr^{2}  \nonumber \\
&&+(r^{2}-n^{2})(d\theta _{1}^{2}+\sin ^{2}\theta _{1}d\phi _{1}^{2})+\alpha
r^{2}(d\theta _{2}^{2}-\sinh ^{2}\theta _{2}dt^{2})  \nonumber \\
ds^{2} &=&F_{E}(r)(d\chi +2n\cos \theta _{1}d\phi
_{1})^{2}+F_{E}^{-1}(r)dr^{2}  \nonumber \\
&&+(r^{2}-n^{2})(d\theta _{1}^{2}+\sin ^{2}\theta _{1}d\phi _{1}^{2})+\alpha
r^{2}(d\theta _{2}^{2}-e^{2\theta _{2}}dt^{2})  \nonumber \\
ds^{2} &=&F_{E}(r)(d\chi +2n\cos \theta _{1}d\phi
_{1})^{2}+F_{E}^{-1}(r)dr^{2}  \nonumber \\
&&+(r^{2}-n^{2})(d\theta _{1}^{2}+\sin ^{2}\theta _{1}d\phi _{1}^{2})+\alpha
r^{2}(-dt^{2}+\sin ^{2}td\phi _{2}^{2})
\end{eqnarray}%
with the same expressions for $F_{E}(r)$ and $\beta $ as in the Euclidean
version.

\section*{Appendix B: T-duality in six dimensions}

The Lagrangian in $D=6$ obtained by dimensional reduction of Type IIB on a
torus and after performing a consistent truncation is given by \cite{Duff2}: 
\begin{eqnarray}
\mathcal{L}_{6B}&=&eR-\frac{1}{2}e(\partial\varphi_1)^2-\frac{1}{2}%
e(\partial\varphi_2)^2-\frac{1}{2}ee^{2\varphi_1}(\partial\chi_1)^2-\frac{1}{%
2}ee^{2\varphi_2}(\partial\chi_2)^2  \nonumber \\
&& -\frac{1}{12}ee^{-\varphi_1-\varphi_2}(F^{NS}_{(3)})^2-\frac{1}{12}%
ee^{\varphi_1-\varphi_2}(F^{RR}_{(3)})^2+\chi_2dA^{NS}_{(2)}\wedge
dA^{RR}_{(2)}  \label{6IIB}
\end{eqnarray}
where $F_{(3)}^{NS}=dA^{NS}_{(2)}$ and $F^{RR}_{(3)}=dA^{RR}_{(2)}+%
\chi_1dA_{(2)}^{NS}$. This Lagrangian is related by T-duality in $D=5$ to a
different six-dimensional theory obtained by making a consistent truncation
of Type IIA compactified on a four-dimensional torus. The corresponding
Lagrangian is given by: 
\begin{eqnarray}
\mathcal{L}_{6A}&=&eR-\frac{1}{2}e(\partial\varphi_1)^2-\frac{1}{2}%
e(\partial\varphi_2)^2 -\frac{1}{48}ee^{\frac{\varphi_1}{2}-\frac{3\varphi_2%
}{2}}(F_{(4)})^2 -\frac{1}{12}ee^{-\varphi_1-\varphi_2}(F_{(3)})^2  \nonumber
\\
&&-\frac{1}{4}ee^{\frac{3\varphi_1}{2}-\frac{\varphi_2}{2}}(F_{(2)})^2
\label{6IIA}
\end{eqnarray}
where $F_{(4)}=dA_{(3)}-dA_{(2)}\wedge A_{(1)}$, $F_{(3)}=dA_{(2)}$
corresponds to the NS-NS 3-form $F_{(3)1}$ and $F_{(2)}=dA_{(1)}$ is the RR
2-form $\mathcal{F}^1_{(2)}$, with the index `1' denoting here the first
reduction step from $D=11$ to $D=10$.

Let us focus on Type IIA theory first. Under a dimensional reduction using
the formulae from the previous appendix we have: 
\begin{eqnarray}
ds_{6}^{2}&=&e^{\frac{\varphi}{\sqrt{6}}}ds_{5}^2+e^{\frac{-3\varphi}{\sqrt{6%
}}}(dz+\mathcal{A}_{(1)})^2
\end{eqnarray}
and we obtain the following 5-dimensional Lagrangian: 
\begin{eqnarray}
\mathcal{L}_{5A}&=&eR-\frac{1}{2}e(\partial\varphi_1)^2-\frac{1}{2}%
e(\partial\varphi_2)^2-\frac{1}{2}e(\partial\varphi)^2-\frac{1}{48}ee^{-%
\frac{3\varphi}{\sqrt{6}}+\frac{\varphi_1}{2}-\frac{3\varphi_2}{2}%
}(F_{(4)}^{\prime})^2  -\frac{1}{12}ee^{\frac{\varphi}{\sqrt{6}}+\frac{\varphi_1}{2}-\frac{3\varphi_2}{2}}(F_{(3)1})^2\nonumber \\
&&-\frac{1}{12}ee^{-\frac{2\varphi}{\sqrt{6}}%
-\varphi_1-\varphi_2}(F_{(3)}^{\prime})^2-\frac{1}{2}ee^{-\frac{4\varphi}{%
\sqrt{6}}}\mathcal{F}_{(2)}^2  \nonumber \\
&&-\frac{1}{4}ee^{-\frac{\varphi}{\sqrt{6}}+\frac{3\varphi_1}{2}-\frac{%
\varphi_2}{2}}(F_{(2)}^{\prime})^2-\frac{1}{4}ee^{\frac{2\varphi}{\sqrt{6}}%
-\varphi_1-\varphi_2}(F_{(2)1})^2-\frac{1}{2}ee^{\frac{3\varphi}{\sqrt{6}}+%
\frac{3\varphi_1}{2}-\frac{\varphi_2}{2}}(d A_{(0)1})^2  \label{5A}
\end{eqnarray}
where the field strengths are defined as follows: 
\begin{eqnarray}
F_{(2)}^{\prime}&=&dA_{(1)}-dA_{(0)1}\wedge\mathcal{A}_{(0)},
~~~~~~~F_{(3)}^{\prime}=dA_{(2)}-dA_{(1)}\wedge\mathcal{A}_{(1)}  \nonumber
\\
F_{(3)1}&=&dA_{(2)1}+dA_{(1)}\wedge A_{(1)}-dA_{(2)}\wedge A_{(0)1}, ~~~~~
F_{(4)}^{\prime}=dA_{(3)}-dA_{(2)}\wedge A_{(1)}-F_{(3)1}\wedge\mathcal{A}%
_{(1)}  \nonumber
\end{eqnarray}
while $\mathcal{F}_{(2)}=d\mathcal{A}_{(1)}$ and $F_{(2)1}=dA_{(1)1}$. Upon
dualising $F_{(4)}$ to a 1-form field strength $d\chi^{\prime}$ its kinetic
term in the above Lagrangian will be replaced by: 
\begin{eqnarray}
-\frac{1}{2}ee^{\frac{3\varphi}{\sqrt{6}}-\frac{\varphi_1}{2}+\frac{%
3\varphi_2}{2}}(d\chi^{\prime})^2+\chi^{\prime}F_{(3)}^{\prime}\wedge
F_{(2)}^{\prime}+\chi^{\prime}F_{(3)1}\wedge\mathcal{F}_{(2)}
\end{eqnarray}
If we perform the field redefinitions: 
\begin{eqnarray}
A_{(1)}^{\prime}&=&A_{(1)}-A_{(0)1}\wedge\mathcal{A}_{(1)}, ~
A_{(2)}^{\prime}=A_{(2)}-A_{(1)1}\wedge\mathcal{A}_{(1)}, ~~
A_{(2)1}^{\prime}=A_{(2)1}+A_{(1)1}\wedge A_{(1)}^{\prime}  \nonumber
\end{eqnarray}
we find: 
\begin{eqnarray}
F_{(2)}^{\prime}=dA_{(1)}^{\prime}+A_{(0)1}\wedge \mathcal{F}_{(2)}, ~~~~~~~
F_{(3)}^{\prime}=dA_{(2)}^{\prime}-A_{(1)1}\wedge\mathcal{F}_{(2)}  \nonumber
\\
F_{(3)1}=dA_{(2)1}^{\prime}+dA_{(1)}^{\prime}\wedge
A_{(1)1}-A_{(0)1}(dA_{(2)}^{\prime}-A_{(1)1}\wedge\mathcal{F}_{(2)}) 
\nonumber \\
\chi^{\prime}F_{(3)}^{\prime}\wedge
F_{(2)}^{\prime}+\chi^{\prime}F_{(3)1}\wedge\mathcal{F}_{(2)}=\chi^{%
\prime}(dA_{(2)}^{\prime}\wedge dA_{(1)}^{\prime}+dA_{(2)1}^{\prime}\wedge%
\mathcal{F}_{(2)})
\end{eqnarray}
Similarly, for the dimensional reduction of Type IIB Lagrangian we obtain: 
\begin{eqnarray}
\mathcal{L}_{5B}&=&eR-\frac{1}{2}e(\partial\varphi_1)^2-\frac{1}{2}%
e(\partial\varphi_2)^2-\frac{1}{2}e(\partial\varphi)^2-\frac{1}{2}%
ee^{2\varphi_1}(\partial\chi_1)^2-\frac{1}{2}ee^{2\varphi_2}(\partial%
\chi_2)^2  \nonumber \\
&&-\frac{1}{12}ee^{-\frac{2\varphi}{\sqrt{6}}+\varphi_1-%
\varphi_2}(F_{(3)}^{^{\prime}~RR})^2-\frac{1}{12}ee^{-\frac{2\varphi}{\sqrt{6%
}}-\varphi_1-\varphi_2}(F_{(3)}^{^{\prime}~NS})^2-\frac{1}{2}ee^{-\frac{%
4\varphi}{\sqrt{6}}}\mathcal{F}_{(2)}^2-\frac{1}{4}ee^{\frac{2\varphi}{\sqrt{%
6}}+\varphi_1-\varphi_2}(F_{(2)1}^{RR})^2  \nonumber \\
&&-\frac{1}{4}ee^{\frac{2\varphi}{\sqrt{6}}-\varphi_1-%
\varphi_2}(F_{(2)1}^{NS})^2-\chi_2dA_{(2)}^{RR}\wedge
dA_{(1)1}^{NS}+\chi_2dA_{(2)}^{NS}\wedge dA_{(1)1}^{RR}  \label{5B}
\end{eqnarray}
where $F_{(2)1}^{NS}=dA_{(1)1}^{NS}$, $\mathcal{F}_{(2)}=d\mathcal{A}_{(1)}$
and: 
\begin{eqnarray}
F_{(3)}^{^{\prime}~NS}=dA_{(2)}^{NS}-dA_{(1)1}^{NS}\wedge\mathcal{A}_{(1)},
~~~~~~~F_{(2)1}^{RR}=dA_{(1)1}^{RR}+\chi_1 dA_{(1)1}^{NS}  \nonumber \\
F_{(3)}^{^{\prime}~RR}=dA_{(2)}^{RR}-dA_{(1)1}^{RR}\wedge \mathcal{A}%
_{(1)}+\chi_1dA_{(2)}^{NS}\wedge\mathcal{A}_{(1)}
\end{eqnarray}
As shown in \cite{Duff2}, the $T$-duality rules relating the two
truncated theories (\ref{5A}) and (\ref{5B}) are: 
\begin{eqnarray}
A_{(0)1}\rightarrow\chi_1, ~~~A_{(1)1}\rightarrow\mathcal{A}_{(1)}, ~~~%
\mathcal{A}_{(1)}\rightarrow A_{(1)1}^{NS},~~~\chi^{\prime}\rightarrow\chi_2,
\nonumber \\
A_{(1)}^{\prime}\rightarrow A_{(1)1}^{RR}, ~~~ A_{(2)}^{\prime}\rightarrow
A_{(2)}^{NS}, ~~~A_{(2)1}^{\prime}\rightarrow - A_{(2)}^{RR}
\end{eqnarray}
together with a rotation of the scalars: 
\begin{eqnarray}
\left(%
\begin{array}{c}
\varphi_1 \\ 
\varphi_2 \\ 
\varphi%
\end{array}%
\right)_{II A,B}&=&\left(%
\begin{array}{ccc}
\frac{3}{4} & -\frac{1}{4} & -\frac{\sqrt{6}}{4} \\ 
-\frac{1}{4} & \frac{3}{4} & -\frac{\sqrt{6}}{4} \\ 
-\frac{\sqrt{6}}{4} & -\frac{\sqrt{6}}{4} & -\frac{1}{2}%
\end{array}%
\right)\left(%
\begin{array}{c}
\varphi_1 \\ 
\varphi_2 \\ 
\varphi%
\end{array}%
\right)_{II B,A}
\end{eqnarray}
which takes care of the dilaton couplings of the field strengths.

\end{document}